\documentclass[pra,twocolumn,floatfix,superscriptaddress,longbibliography,notitlepage]{revtex4-2}
\usepackage[off]{auto-pst-pdf}
\usepackage{caption}
\usepackage{ragged2e}
\DeclareCaptionJustification{justified}{\justifying}
\usepackage{caption}
\usepackage{graphicx}
\usepackage{subfig}
\usepackage{hyperref}
\usepackage[T1]{fontenc}
\usepackage{mathtools}
\usepackage{derivative}
\usepackage{bbold}
\usepackage{float}
\usepackage[thinc]{esdiff}
\usepackage{dcolumn}
\usepackage{bm}
\usepackage{tikz}
\usetikzlibrary{plotmarks}

\usepackage{auto-pst-pdf} 
\usepackage{cleveref}
\usepackage{braket}

\DeclareMathOperator{\Tr}{Tr}
\begin{document}

\title{Navigating the phase diagram of quantum many-body systems in phase space}%


\author{Khadija El Hawary}
\email{khadija\_elhawary@um5.ac.ma}
\affiliation{ESMaR, Faculty of Sciences, Mohammed V University in Rabat, Av. Ibn Battouta, B.P. 1014, Agdal, Rabat, Morocco.}
\author{Mohamed Azzouz}
\affiliation{Al Akhawayn University, School of Science and Engineering, PO Box 104, Hassan II Avenue, 53000 Ifrane, Morocco.}
\author{Morad El Baz}
\affiliation{ESMaR, Faculty of Sciences, Mohammed V University in Rabat, Av. Ibn Battouta, B.P. 1014, Agdal, Rabat, Morocco.}
\author{Sebastian Deffner}
\affiliation{Department of Physics, University of Maryland, Baltimore County, Baltimore, MD 21250, USA}
 \affiliation{National Quantum Laboratory, College Park, MD 20740, USA}
\author{Bartłomiej Gardas}
\author{Zakaria Mzaouali}%
 \email{zmzaouali@iitis.pl}
\affiliation{Institute of Theoretical and Applied Informatics, Polish Academy of Sciences, Bałtycka 5, Gliwice, 44-100, Poland.}
\date{\today}
            
\begin{abstract}
We demonstrate the unique capabilities of the Wigner function, particularly in its positive and negative parts, for exploring the phase diagram of the spin$-(\frac{1}{2\!}-\!\frac{1}{2})$ and spin$-(\frac{1}{2}\!-\!1)$ Ising-Heisenberg chains. We highlight the advantages and limitations of the phase space approach in comparison with the entanglement concurrence in detecting phase boundaries. We establish that the equal angle slice approximation in the phase space is an effective method for capturing the essential features of the phase diagram, but falls short in accurately assessing the negativity of the Wigner function for the homogeneous spin$-(\frac{1}{2}\!-\!\frac{1}{2})$ Ising-Heisenberg chain. In contrast, we find for the inhomogeneous spin$-(\frac{1}{2}\!-\!1)$ chain that an integral over the entire phase space is necessary to accurately capture the phase diagram of the system. This distinction underscores the sensitivity of phase space methods to the homogeneity of the quantum system under consideration.
\end{abstract}

\maketitle

\section{Introduction}
The intricate landscape of quantum systems has continually posed challenges and opportunities for understanding the behavior of matter at its most fundamental level~\cite{Review_matter_nano,quantum_matter_meets_QI,quantum_matter}. Quantum phases of matter, like superconductivity or topological phases, exhibit unique properties that emerge from quantum effects at a microscopic level~\cite{Lee_1973,Sachdev_2011}. These phases are more than mere theoretical interests; they hold the key to overcoming some of the major challenges in advancing the development of quantum computers~\cite{topo_QC,ayral2023quantum,head2020quantum,huang2020superconducting,Wendin_2017,superconducting_information_outlook}. For instance, understanding and harnessing these phases can lead to the development of more stable qubits, which are the fundamental units of quantum computation~\cite{atature2018material}. Qubits in certain quantum phases are less susceptible to decoherence, a major drawback where quantum information gets lost to the environment~\cite{decoherence_qubit,siddiqi2021engineering, Touil_2020, Touil_2021}. Furthermore, exploration in this realm could lead to the discovery of new materials and methods that allow for quantum coherence and entanglement over longer times and distances~\cite{mzaouali2019long,abaach2021pairwise,Abaach_2021}, significantly enhancing computational power and efficiency~\cite{Deffner_2021,smierzchalski_2024}. This synergy between the study of quantum phases of matter and quantum computing paves the way for revolutionary advancements in computing, encryption, and information processing~\cite{quantum_tech_2nd_rev}.

Along this line, low-dimensional magnetic materials, such as one- or two-dimensional systems, have garnered significant attention in the field of quantum computation due to their unique physical properties~\cite{computing_spin}. These materials often exhibit strong quantum fluctuations and reduced symmetry, which can lead to exotic magnetic states like quantum spin liquids and topological order, which are robust for quantum information processing~\cite{rev_quantum_spin_liquid_2016,rev_quantum_spin_liquid_2017,2023_Sonnenschein_PRB}. Of particular interest are diamond-type chains which can be described using Ising-like or quantum anisotropic Heisenberg models~\cite{okamoto_2002,streka_2004,streka_2006,valverde2008phase} and are exactly solvable using the decorated transformation method introduced by Fisher~\cite{fisher1959transformations}. 
Experimentally, diamond chains can describe and capture the magnetic properties of minerals composed primarily of copper carbonate hydroxide, such as natural azurite $\text{Cu}_{3} (\text{CO}_{3})_{2}(\text{OH})_{2}$~\cite{honecker2001frustrated, kikuchi2003magnetic,kikuchi2004experimental,kikuchi2005experimental,kikuchi2005magnetic,ohta2003high,okubo2004submillimeter,ohta2004recent,okubo2004high,guillou2002first}. The copper ions in azurite can act as magnetic spins that interact with each other in a way that closely approximates the interactions described by the Heisenberg diamond chain model~\cite{kikuchi2003magnetic,kikuchi2004experimental,kikuchi2005experimental,kikuchi2005magnetic, ohta2003high,okubo2004submillimeter,ohta2004recent,okubo2004high,guillou2002first}. This makes diamond chains a valuable natural testbed for studying the properties and behaviors of azurite, such as quantum phase transitions~\cite{honecker2001frustrated,jeschke2011multistep,rojas2011exactly,carvalho2019correlation,ZHENG2022126444}, and magnetization plateaus~\cite{pereira2008magnetization,pereira2009magnetocaloric,ananikian2012magnetic, lisnii2011distorted}.

The rich and complex phase diagram as well as the simplicity and solvability of the diamond Ising-Heisenberg structures make them strong candidates for synthesising new materials with tailor-made magnetic properties for quantum computation~\cite{quantum_annealing_antiferro,quantum_simulation_antiferro,quantum_computing_antiferro_spin_cluster,quantum_computing_antiferro_molecular_eng,ferro_quantum_computing}. 
\begin{figure*}
    \centering
    \subfloat[The ATIH lattice. \label{lattice}]{\includegraphics[width=0.5\textwidth]{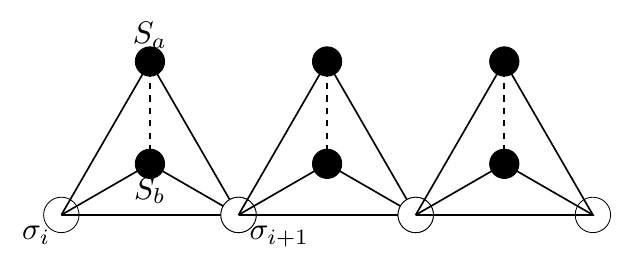}}%
    \subfloat[Zoom on single unit cell of the ATIH model, and its effective mapping to an Ising model via DIT. \label{lattice_dit}]{\includegraphics[width=0.5\textwidth]{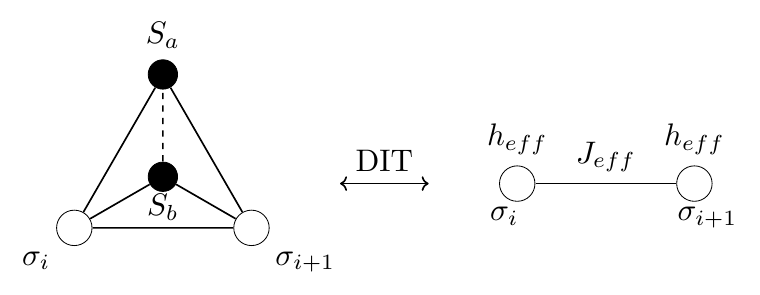}}
    \caption{The asymmetric tetrahedral Ising-Heisenberg (ATIH) chain as outlined by the Hamiltonian, Eq.~\eqref{ham}. (a) Shows three unit cells of the lattice, where white vertices symbolize the sites connected via Ising interactions, whereas black vertices represent those interacting via Heisenberg interactions, indicated by $J$ for Ising and $J_{\alpha}$ ($\alpha = x, y, z$) for Heisenberg connections, respectively.  (b) Single unit cell of the ATIH chain and its effective mapping into an Ising model. Fisher's decoration-iteration transformation (DIT), changes the Heisenberg edge within the tetrahedral structure into ``decorative'' elements~\cite{fisher1925theory}. This process essentially replaces the Heisenberg interactions with an equivalent Ising-type interaction. The modification preserves the effect of the Heisenberg spins on the system's magnetic behaviors within a simpler analytical framework. The resulting effective Ising model underlies the core physical phenomena of the original configuration into a form that is significantly more tractable for examining phase transitions and critical phenomena. Appendix~\eqref{ATIH_thermo} elaborates on the statistical and thermodynamics of the ATIH chain, achieved through the DIT method.}
    \label{fig_lattice}
\end{figure*}
Quantum information theory provides the necessary tools and framework for a succinct analysis of the potential of diamond chains for applications in quantum technologies~\cite{nielsen2002quantum}. For the diamond Ising-Heisenberg chain, the phase diagram has been analyzed through quantum entanglement~\cite{rojas2012thermal,torrico2014pairwise,ABGARYAN20155,ananikian2012magnetic,ROJAS2017506,Qiao_2015,Kuzmak_2023,Freitas_2019,Chakhmakhchyan_2012,ROJAS2017367,carvalho2018quantum}, and general forms of quantum correlations quantified via quantum discord~\cite{GAO201510,Zheng_2018,cheng2017finite,FAIZI2014251} and quantum Fisher information~\cite{REN2022105542}. However, an analysis with a more intuitive understanding of quantum phenomena, akin to classical mechanics, while still capturing the intricacies of quantum behavior is lacking. The Wigner function does just that by enabling the visualization of quantum features like superposition and entanglement, through its negative part in phase space, thereby offering a different angle to analyze and understand the properties inherent in quantum many-body systems~\cite{wigner_book}.

The rapid development of quantum technologies in the 21st century allowed for the possibility of experimental measurement of the phase space, which pushed for the use of phase space techniques for analyzing the critical properties of quantum-many body systems~\cite{rundle_review}.  Recently, it has been established that the Wigner function is a bonafide measure of first-, second-, and infinite-order quantum phase transitions in the Ising and Heisenberg chains~\cite{mzaouali2019discrete,MILLEN2023169459}. In this paper, we build on these studies and we focus on the role of the Wigner function and its negative part in delimiting the phase diagram of an exotic quantum spin chain, i.e. the homogeneous spin$-(\frac{1}{2\!}-\!\frac{1}{2})$ and inhomogeneous spin$-(\frac{1}{2}\!-\!1)$ Ising-Heisenberg chains. The quasi-probability nature of the Wigner function in phase space, allows us to identify classical and quantum regimes, offering a unique perspective on quantum states and their critical properties, which is pivotal for the development of quantum technologies~\cite{Ferry_APR_2018}.

The relevance of our study lies in its comparison with other quantum information tools, notably the concurrence which is a measure of entanglement and has been instrumental in detecting phase boundaries in quantum systems~\cite{Amico_Review_Entanglement,Grzegorz_Review_Entanglement}. Our results highlight the advantages and limitations of using phase space methods in contrast to the entanglement concurrence. This comparison is vital in understanding the most effective approaches for studying quantum many-body systems, as different methods can yield varying insights into the critical properties of these systems. Furthermore, we shed light on the concept of the equal angle slice approximation in phase space. We show its effectiveness in capturing the essential features of the phase diagram, particularly for the homogeneous spin$-(\frac{1}{2\!}-\!\frac{1}{2})$ Ising-Heisenberg chain. Our investigation into the Wigner function's negativity provides a profound understanding of the quantum states in these chains. In contrast, for the inhomogeneous spin$-(\frac{1}{2\!}-\!1)$ chain, we find that a more comprehensive approach, involving an integral over the entire phase space, is necessary. This distinction underscores the sensitivity of phase space methods to the homogeneity of the quantum system under consideration, highlighting the need for tailored approaches in studying different quantum systems.
\section{The asymmetric tetrahedron Ising-Heisenberg chain }
\begin{figure*}
    \centering
    \subfloat[\label{phase_diagram_1}Spin$-(\frac{1}{2}\!-\!\frac{1}{2})$]
    {\includegraphics[width=0.5\textwidth]{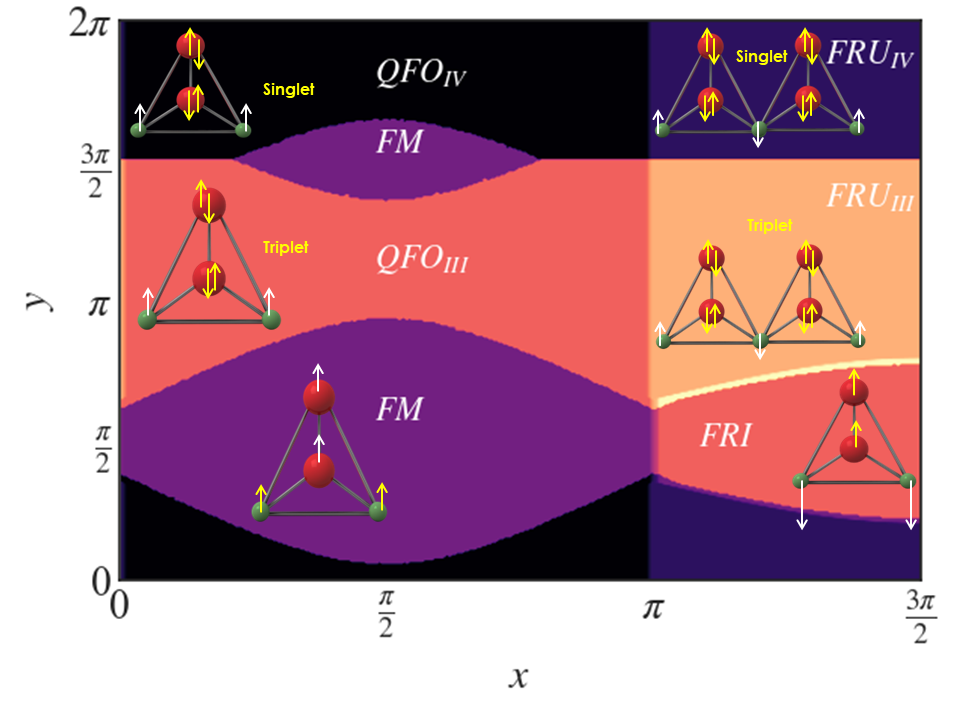}}%
    \subfloat[\label{phase_diagram_2}Spin$-(\frac{1}{2}\!-\!1)$]{\includegraphics[width=0.5\textwidth]{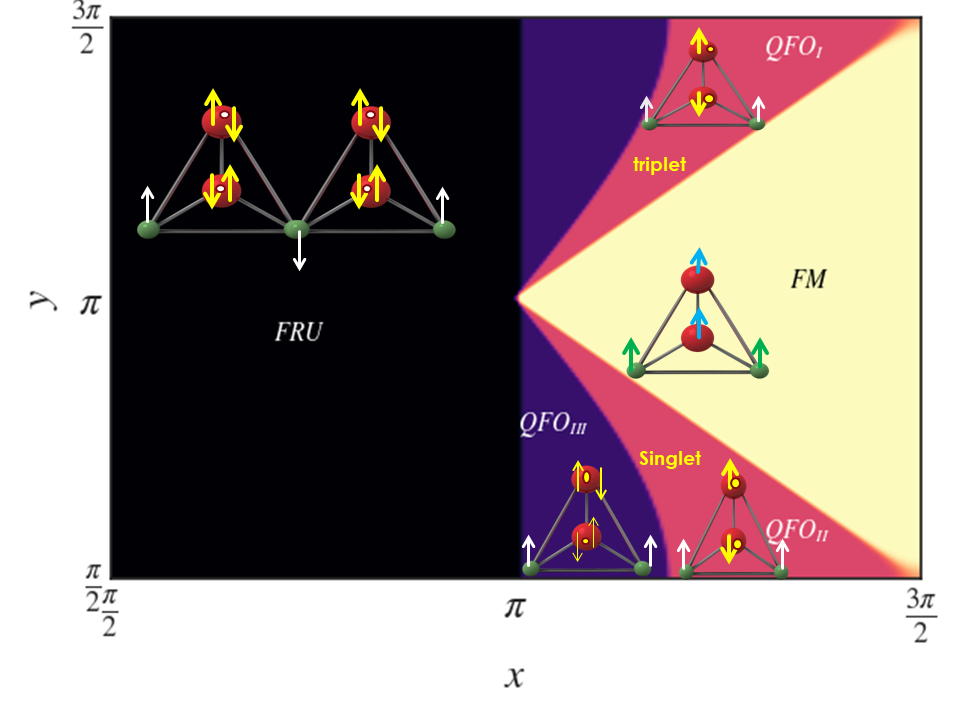}}
    \caption{Phase diagram for the asymmetric tetrahedral Ising-Heisenberg chain, Eq.~\eqref{ham}, for two different spin configurations. Panel (a) shows the phase diagram for the spin$-(\frac{1}{2}\!-\!\frac{1}{2})$ case, utilizing parameters from Eq.~\eqref{phase_diagram_1_parameters}. The spin alignments are represented by arrows, with up or down arrows indicating the two possible spin states at the Ising and Heisenberg nodes. Panel (b) illustrates the phase diagram for the spin$-(\frac{1}{2}\!-\!1)$ case, using parameters from Eq.~\eqref{phase_diagram_2_parameters}. Here, up and down arrows denote the Ising nodes, while a combination of up, dot, and down illustrates the spin-1 configurations on the Heisenberg edges. Magnetic phases are labeled as FM (ferromagnetic), QFO (quantum ferromagnetic), FRI (ferrimagnetic), and FRU (frustrated ferromagnetism). The FM state is a product state, while the QFO states are entangled and differ due to the distinct orientation probabilities of spins in the Heisenberg edges. The FRI state emerge from the superposition of ferromagnetic and antiferromagnetic state, causing a competition in the spin alignment.  The FRU states exhibit non-classical behaviour caused by competing interactions, leading to frustration where classical ordering is absent, suggesting a higher degree of quantum correlation.
    Appendix~\eqref{spect_ATIH} elaborates on the model's spectrum and the specific states of each phase for both scenarios.}
    \label{phase_diagram}
\end{figure*}
The diamond chain described via the Ising-Heisenberg model comprises a combination of Ising spins located at nodes and interstitial anisotropic Heisenberg spins. The single unit cell of the chain is sketched in Fig.~\eqref{lattice}. The Hamiltonian operator is formulated as follows:
\begin{align}
\label{ham}
H =& \sum_{i=1}^N H_{i,i+1} \nonumber \\ =&\sum_{i=1}^N J(S^z_{a,i} + S^z_{b,i})(\sigma_i  + \sigma_{i+1}) 
 + J\sigma_i  \sigma_{i+1}  \nonumber \\ & + J_{x}S^x_{a,i} S^x_{b,i} 
+  J_{y}S^y_{a,i} S^y_{b,i}+  J_{z}S^z_{a,i} S^z_{b,i} \nonumber \\ & + \frac{h_0}{2}(\sigma_i  + \sigma_{i+1}) + h( S^z_{a,i} + S^z_{b,i}) ,
\end{align}
where $N$ is the number of cells, ${(S}_{a,i} , {S}_{b,i})$ are the Heisenberg interstitial spins interacting via $J_{\alpha}, \alpha=(x, y, z)$, and $\sigma_i$'s are Ising spins interacting via $J$. The longitudinal external magnetic field $h$ ($h_0$) operates on Heisenberg (Ising) spins. The Hamiltonian, Eq.~\eqref{ham}, is symmetric under the exchange of the Ising spins, i.e. $\sigma_i \leftrightarrow \sigma_{i+1}$, and Heisenberg spins, that is $S_a^z \leftrightarrow S_b^z$. Additionally, 
the asymmetric tetrahedron Ising-Heisenberg (ATIH) chain, Eq.~\eqref{ham}, is invariant under internal spin symmetry, i.e. $H(\sigma_i, \sigma_{i+1})=H(-\sigma_i,-\sigma_{i+1})$. To this end, we consider two cases where the Heisenberg spin takes $S=\frac{1}{2}$ and $S=1$. 

The ATIH chain is an exactly solvable model by making use of the decoration-iteration transformation~(DIT) introduced by Fisher~\cite{fisher1959transformations}. The DIT or star-triangle transformation involves replacing parts of a lattice model with simpler, equivalent structures without altering the physical properties or critical behavior of the system. This is achieved by ``decorating'' the lattice with additional sites or interactions in a way that allows for an exact transformation of the partition function, which describes the statistical properties of the system~\cite{Lee_1973}. The transformed model is easier to analyze or solve, allowing to gain insights into the original, more complex system. For the ATIH chain, the DIT maps it to an effective Ising chain, as described in Fig.~\eqref{lattice_dit}.

In the following, we restrict our analysis to the case where $J_x=J_y$, which reduces the Heisenberg edge to an $XXZ$ type interaction. 
Fig.~\eqref{phase_diagram_1} shows the phase diagram of the ATIH spin-($\frac{1}{2}-\frac{1}{2}$) chain under the following new set of parameters
\begin{align}
    J&=-\sin{(x)}, & J_z&=-\sin{(y)}, & J_x&= 2\cos{(y)},
    \label{phase_diagram_1_parameters}
\end{align}
which restrict the system in a region with competing interaction parameters, i.e. $|J|\leq 1, |J_z|\leq 1 $, and $|J_{x}|\leq 2$,
leading to several ground state energies. Outside this region there are no new phases~\cite{valverde2008phase}. 

For the ATIH spin-($\frac{1}{2}-1$) chain we use the following parameters
\begin{align}
    J&=J_z=\sin{(x)},& J_x&=J_y=2\sin{(y)},
    \label{phase_diagram_2_parameters}
\end{align}
to draw the phase diagram represented in Fig.~\eqref{phase_diagram_2}.  This re-parametrization essentially maps the original interaction parameters $J$, $J_z$, $J_+$, and $J_-$ into a new coordinate system defined by $x$ and $y$, facilitating a more tractable exploration of the model's behavior across different parameter regimes. The mapping reveals critical points where multiple phases converge and provides a clearer understanding of the phase transitions within the ATIH chain~\cite{valverde2008phase}. 
\section{Figures of Merit}
We analyze the phase diagram of the ATIH chain, Eq.~\eqref{ham}, using tools from quantum information theory, such as entanglement measures and phase space methods, i.e. the Wigner function. The density matrix of a single cell $\rho_{i,a,b,i+1}$, comprised by two Ising nodes located at sites ($i, i+1$) and two Heisenberg nodes ($a,b$), described by the ATIH model, Eq.~\eqref{ham}, can be written as:
\begin{equation}
\label{rho_total}
    \rho_{i,a,b,i+1}\!=\!\frac{1}{d}\sum_{\alpha ,\eta= 0}^{1} \sum_{\beta ,\gamma= 0}^{3} \langle \sigma^{\alpha}_i S^\beta_{a}S^\gamma_{b}\sigma^\eta_{i+1}\rangle \sigma^\alpha_i \otimes S^\beta_{a}\otimes S^\gamma_{b}\otimes \sigma^\eta_{i+1},
\end{equation}
with $d$ being the dimension of the single cell's Hilbert space, $\sigma_i^0$ denoting the identity matrix and $\sigma_i^1 = \sigma^z$  at site $i$ (and site $i+1$). In the same fashion, $S$ represent the spin operator with $S^0 = \mathbb{1}, S^1 = S^x, S^2 = S^y,$ and $S^3 = S^z$. For the homogeneous case, i.e. spin$-(\frac{1}{2} , \frac{1}{2}) $, $d= 16$ and for the  spin$-(\frac{1}{2} , 1 )$ inhomogeneous chain $d=36$. The averages in the spin-spin correlation functions is taken over the ground state of the ATIH chain, Eq.~\eqref{ham}, and they can be calculated using the transfer matrix approach outlined in Appendix~\eqref{ATIH_thermo}. The detailed expression of the density matrix, Eq.~\eqref{rho_total}, is provided in Appendix~\eqref{dm}, from which we can compute various information-theoretic quantities, such as the entanglement measures and the Wigner function.
\paragraph*{\bf Lower bound concurrence.}
Quantifying the entanglement in quantum systems is a complex task and remains an active area of research in quantum information science~\cite{Grzegorz_Review_Entanglement}. For bipartite qubit systems, analytical formulas for measuring the entanglement are well known, i.e. entanglement of formation, and Wootters concurrence~\cite{wootters1998entanglement}. However, the situation is delicate for higher dimensional quantum systems with multi-parties,  for which the single cell of the ATIH chain, Eq.~\eqref{ham}, falls into. The entanglement in this case is estimated through lower bounds. 

For an $N$-partite quantum state $\rho$, residing in a composite Hilbert space represented by multiple tensor products of $\mathcal{H}$, i.e. $\rho \in \mathcal{H} \otimes \mathcal{H} \otimes ... \otimes \mathcal{H}$,  a foundational lower bound for the concurrence has been established~\cite{zhu2012lower, espoukeh2015lower}. It is expressed as
\begin{equation}
C(\rho) \geq \tau_{N}(\rho) = \sqrt{\frac{d}{2m(d-1)}\sum_{p}\sum_{\alpha \beta} \left(C^{p}_{\alpha\beta}(\rho)\right)^2},
\label{LBCC}
\end{equation}
where the term $m = 2^{N-1} - 1$ denotes the number of bipartitions possible within an $N$-partite system. $\tau_{N}(\rho)$ encodes this lower limit for $C(\rho)$, encapsulating the summation over all bipartite splits indicated by the indices $\alpha, \beta$. Here, $C^{p}_{\alpha\beta}(\rho)$ measures the entanglement across a partition $p$, assuming a uniform dimension $d$ across each segment.

The individual concurrences $C^{p}_{\alpha \beta}$ are determined through a widely recognized formula~\cite{wootters1998entanglement}:
\begin{equation}
C^{p}_{\alpha \beta} = \max{(0 , \lambda_1 - \lambda_2 -\lambda_3 - \lambda_4)},
\end{equation}
where $\lambda_i$ represents the square roots of the top four eigenvalues, sorted in descending order, of the matrix $R =\rho^{p}_{\alpha \beta}\tilde{\rho}^{p}_{\alpha \beta} $. The modified state $\tilde{\rho}^{p}_{\alpha \beta}$ is defined as $\tilde{\rho}^{p}_{\alpha \beta} =	(S^{p}_{\alpha \beta}\rho^*S^{p}_{\alpha \beta}) $, where $S^{p}_{\alpha \beta} = L_{\alpha}\otimes L_{\beta}$, with $L_{\alpha}$ and $L_{\beta}$ are the generators of the special orthogonal group $SO(d)$.

The peculiarity of the lower bound $\tau_{N}(\rho)$, Eq.~\eqref{LBCC}, is its connection with separability. When $\tau_{N}(\rho)=0$, the quantum state $\rho$ is fully separable. Thus, for non-zero values of $\tau_{N}(\rho)$ some entanglement is present in the system. The lower bound concurrence, Eq.~\eqref{LBCC}, has been useful in several scenarios, such as studying the statics and dynamics of multipartite entanglement in qubit and ququart systems~\cite{Abaach_2021, abaach2021pairwise}, identification of quantum resource for quantum teleportation~\cite{abaach2023long}, as well as studying the performance of quantum heat engines~\cite{el2023performance}.
\paragraph*{\bf Phase space techniques.}
The Wigner function initially served to depict quantum states in phase space with continuous variables. However, for discrete systems, numerous methods have been devoloped to map these quantum systems onto a phase space framework within a discrete-dimensional Hilbert space. In this context, we adopt the approach proposed by Tilma {\it et al}~\cite{tilma2016wigner}, which extends the Wigner function's applicability to arbitrary quantum states. According to this framework, the Wigner function is formulated using the displacement operator $\hat{D}$ and the parity $\hat{\Pi}$ operators as
\begin{equation}
W_{\hat{\rho}}(\Omega)=\left( \frac{1}{\pi \hbar} \right)^n \text{Tr }{ \left(\hat{\rho} \hat{D}(\Omega) \hat{\Pi} \hat{D}^{\dagger}(\Omega) \right)},
\label{eq4.1}
\end{equation}
where $\hat{D}(\Omega) \hat{\Pi} \hat{D}^{\dagger}(\Omega)=\hat{\Delta}(\Omega)$ forms the kernel of this function, $ \hat{\rho} $ is the density matrix describing the system and $\Omega$ represents a complete parametrization of the phase space such that $\hat{D}$ and $\hat{\Pi}$ are defined in terms of coherent states $\hat{D}(\Omega)\ket{0}=\ket{\Omega}$ and $\hat{\Pi}\ket{\Omega}=-\ket{\Omega}$. A distribution $W_{\hat{\rho}}(\Omega)$ can describe a Wigner function over a phase space parametrized by a set of $\Omega$'s, if there exists a kernel $\hat{\Delta}(\Omega)$ that generates $W_{\hat{\rho}}(\Omega)$ according to the Weyl rule
\begin{equation}
	W_{\hat{\rho}}(\Omega)=\text{Tr} \left( \hat{\rho} \hat{\Delta}(\Omega) \right),
	\label{wigner_arbirary_system}
\end{equation}
and, as stated in~\cite{tilma2016wigner}, also satisfy the Stratonovich-Weyl correspondences, which articulate several foundational properties of the Wigner function. Primarily, they allow for a bi-directional reconstruction between the density matrix $\hat{\rho}$ and its representation $W_{\hat{\rho}}(\Omega)$ through a precise mathematical mapping. This mapping not only facilitates the transition from $\hat{\rho}$ to  $W_{\hat{\rho}}(\Omega)$ via the trace formula $W_{\hat{\rho}}(\Omega)=\text{Tr} \left( \hat{\rho} \hat{\Delta}(\Omega) \right)$  but also enables the reconstruction of $\hat{\rho}$ from $W_{\hat{\rho}}(\Omega)$ by integrating over $\Omega$ with the kernel $\hat{\Delta}(\Omega)$. Another critical aspect of these correspondences is the reality and normalization to unity of $W_{\hat{\rho}}$, ensuring that it always represents a valid quantum state. Additionally, the invariance of  $\hat{\rho}$ under global unitary operations implies a corresponding invariance in $W_{\hat{\rho}}$, preserving the physical characteristics of the quantum state within the phase space. Lastly, a distinctive feature of the Wigner function is its ability to quantify the overlap between two states $\hat{\rho}^{\prime}$ and $\hat{\rho}^{\prime\prime}$. This is executed through a definite integral over $\Omega$ via $\int_{\Omega} W_{\hat{\rho}^{\prime}} W_{\hat{\rho}^{\prime\prime}} d\Omega=\text{Tr} \left( \hat{\rho}^{\prime} \hat{\rho}^{\prime\prime} \right)$ which highlights a unique property that sets the Wigner function apart in the analysis of quantum states.

An extension of Eq.~\eqref{wigner_arbirary_system} to finite-dimensional systems requires the construction of a kernel $\hat{\Delta}(\Omega)$ that reflects the symmetries of the system at hand. For spin systems, Tilma \textit{et al}~\cite{tilma2016wigner} argued that their Wigner functions can be generated via a spin$-j$ representation of SU(2) under the following kernels:
\begin{equation}
    \hat{\Delta}^{[d]}(\Omega)=\frac{1}{d}\left[\hat{U}(\Omega) \hat{\Pi}^{[d]} \left( \hat{U}(\Omega) \right)^{\dagger}  \right],
    \label{wigner_kernel_arbitrary_system}
\end{equation}
with
\begin{equation}
    \hat{\Pi}^{[d]}=\openone_{d\times d} - \mathcal{N}(d) \hat{\zeta}_{d\times d},
\end{equation}
where $d\!=\!2j+1$ is the dimension of the Hilbert space with $j$ being the spin number, $\mathcal{N}(d)\!=\!\sqrt{d(d+1)(d-1)/2}$, and $\hat{\zeta}$ is a diagonal matrix with entries $\sqrt{\frac{2}{d(d-1)}}$ except the last element $\hat{\zeta}_{d,d} = -\sqrt{\frac{2(d-1)}{d}}$.
In this case, the operator $\hat{U}(\Omega)$ represents the SU(2) rotations under three angles $\theta, \varphi,$ and $\phi$:
\begin{equation}
    \hat{U}(\Omega)=\hat{U}(\theta,\varphi,\phi)=e^{i\hat{J}_3\varphi}e^{i\hat{J}_2\theta}e^{i\hat{J}_3\phi},
\end{equation}
where the $J_i$'s are the generators of the $d-$dimensional representation of SU(2). For $k-$partite quantum systems, the kernel, Eq.~\eqref{wigner_kernel_arbitrary_system}, extends as
\onecolumngrid
\begin{equation}
   \hat{\Delta}^{[d^k]}  \Bigl\{(\theta_i,\varphi_i)\Bigr\} = \frac{1}{d^k} \Bigl\{ \bigotimes_i^k \hat{U}^{[d]}(\theta_i,\varphi_i)\Bigr\} \hat{\Pi}^{[d^k]}  \Bigl\{ \bigotimes_i^k  \left( \hat{U}^{[d]}(\theta_i,\varphi_i) \right)^{\dagger} \Bigr\}, 
\end{equation}
\twocolumngrid
\noindent which becomes hard to visualize, as the number of angles $\{ \theta_i, \varphi_i \}$ scales with the number of subsystems. Therefore, we work within the equal angle slice approximation, by setting $\theta_i =\theta$ and $\varphi_i = \varphi$, which has been argued to capture the salient properties of quantum states~\cite{tilma2016wigner}. Therefore, under the equal angle slice approximation, the total Wigner function of the ATIH chain, Eq.~\eqref{ham}, in a single cell described by the state $\rho_{i,a,b,i+1}$, Eq.~\eqref{rho_total}, is written as:
\onecolumngrid
\begin{align}
    &W_{\text{cell}}^{\frac{1}{2}-\frac{1}{2}} (\theta,\varphi)= \Tr\left( \rho_{i,a,b,i+1} \hat{\Delta}^{[16]}  (\theta,\varphi)\right), \label{total_wigner_spin_half} \\
    &\hat{\Delta}^{[16]}  (\theta,\varphi)= \frac{1}{16} \Bigl\{ \bigotimes_i^4 \hat{U}^{[2]}(\theta,\varphi)\Bigr\} \hat{\Pi}^{[2^4]}  \Bigl\{ \bigotimes_i^4  \left( \hat{U}^{[2]}(\theta,\varphi) \right)^{\dagger} \Bigr\}, \nonumber \\
    &W_{\text{cell}}^{\frac{1}{2}-1} (\theta,\varphi) = \Tr\left( \rho_{i,a,b,i+1} \hat{\Delta}^{[36]}  (\theta,\varphi)\right), \label{total_wigner_spin_one} \\
    &\hat{\Delta}^{[36]}  (\theta,\varphi) = \frac{1}{36} \left( \hat{\Delta}^{[2]}(\theta,\varphi) \otimes \hat{\Delta}^{[3]}(\theta,\varphi) \otimes \hat{\Delta}^{[3]}(\theta,\varphi) \otimes \hat{\Delta}^{[2]}(\theta,\varphi) \right). \nonumber
\end{align}
\twocolumngrid
The Wigner function, Eq.~\eqref{wigner_arbirary_system}, may take negative values due to its quasi-probability nature. The negativity of the Wigner function, under the equal angle slice approximation,  is defined as:
\begin{equation}
    \mathcal{N}_{W}= \frac{1}{2} \int_{\Omega} \Big( |W_{\rho}(\Omega)| - W_{\rho}(\Omega) \Big) d\Omega.
    \label{negativity_wigner}
\end{equation}
where $d\Omega = \frac{1}{\pi} \sin(2\theta) d\theta d\phi$.
This negativity has a profound physical interpretation as it is interpreted as an indicator of non-classical behavior, distinguishing quantum phenomena from their classical counterparts. In the context of quantum entanglement and quantum correlations, the negativity of the Wigner function is a valuable tool for detecting and quantifying \textit{some} entangled states~\cite{Anatole_Kenfack_2004,siyouri2016negativity}. In what follows,~\Cref{wigner_arbirary_system,LBCC,negativity_wigner} will be our main figures of merit.
\section{Results and discussions}
\paragraph*{Homogeneous ATIH chain. }
\begin{figure*}
    \centering
    \subfloat[Average value of the Wigner function, Eq.~\eqref{wigner_arbirary_system}. \label{average_wigner_4_spin_half}]{\includegraphics[width=0.23\textwidth]{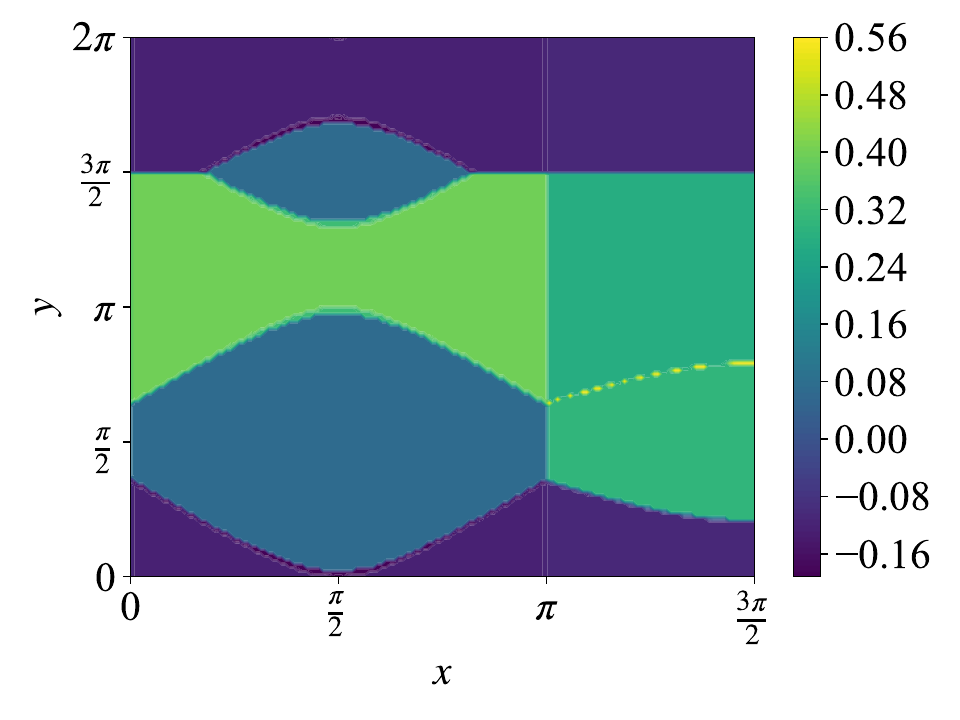}}\quad %
    \subfloat[Negativity of the Wigner function, Eq.~\eqref{negativity_wigner}. \label{negativity_wigner_4_spin_half}]{\includegraphics[width=0.23\textwidth]{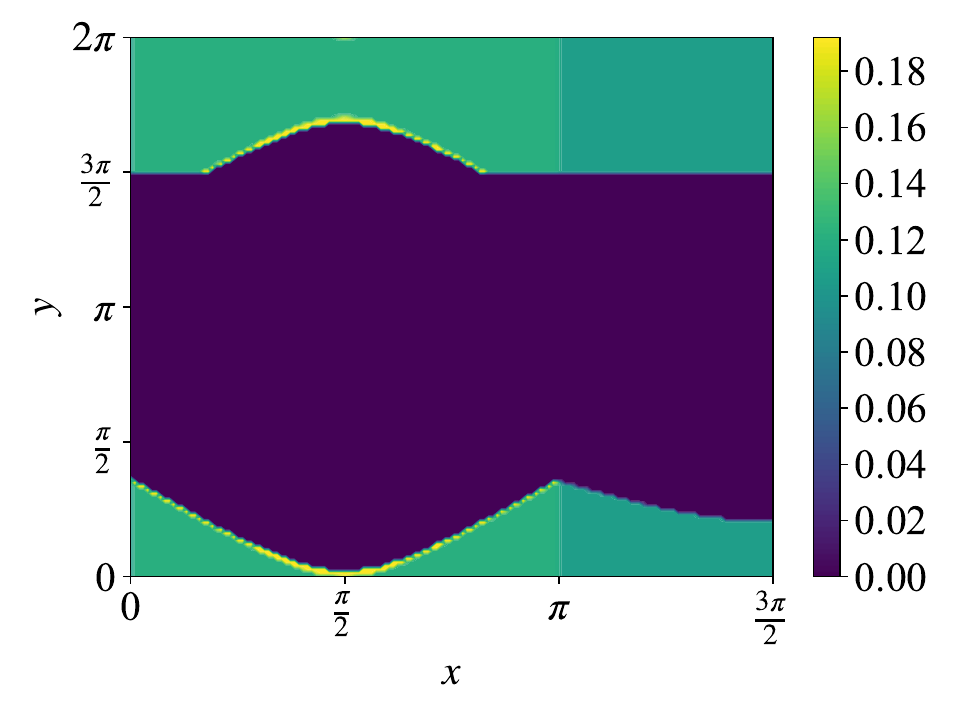}}\quad %
    \subfloat[Negativity of the Wigner function,  Eq.~\eqref{negativity_wigner}, using MC-Integration\label{negativity_wigner_4_spin_half_MC}]{\includegraphics[width=0.23\textwidth]{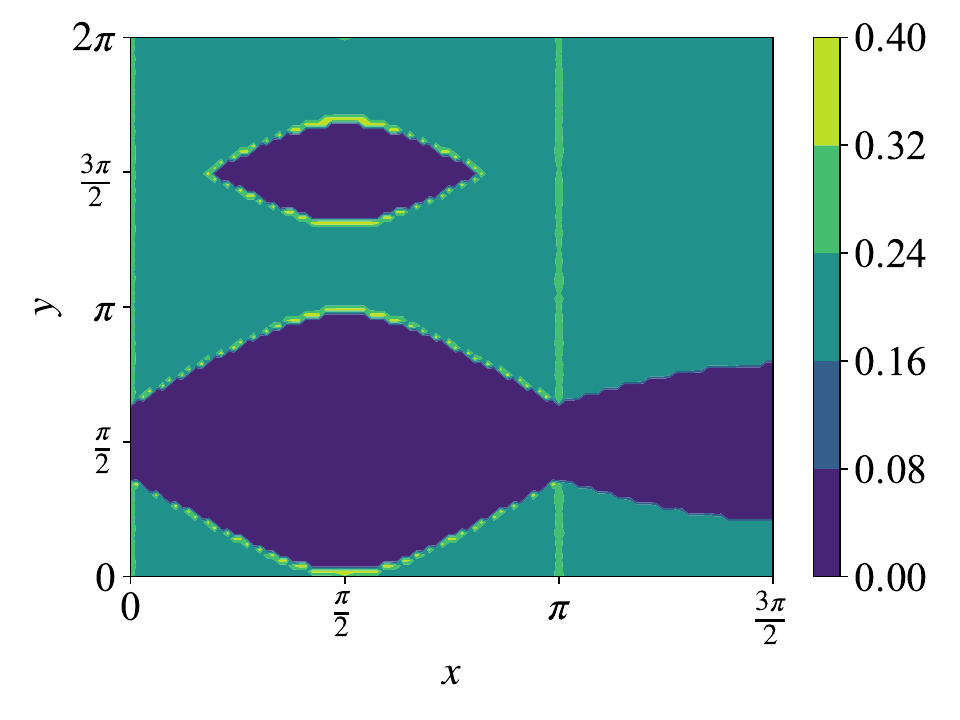}}\quad%
    \subfloat[Lower bound concurrence, Eq.~\eqref{LBCC}. \label{lbc_wigner_4_spin_half}]{\includegraphics[width=0.23\textwidth]{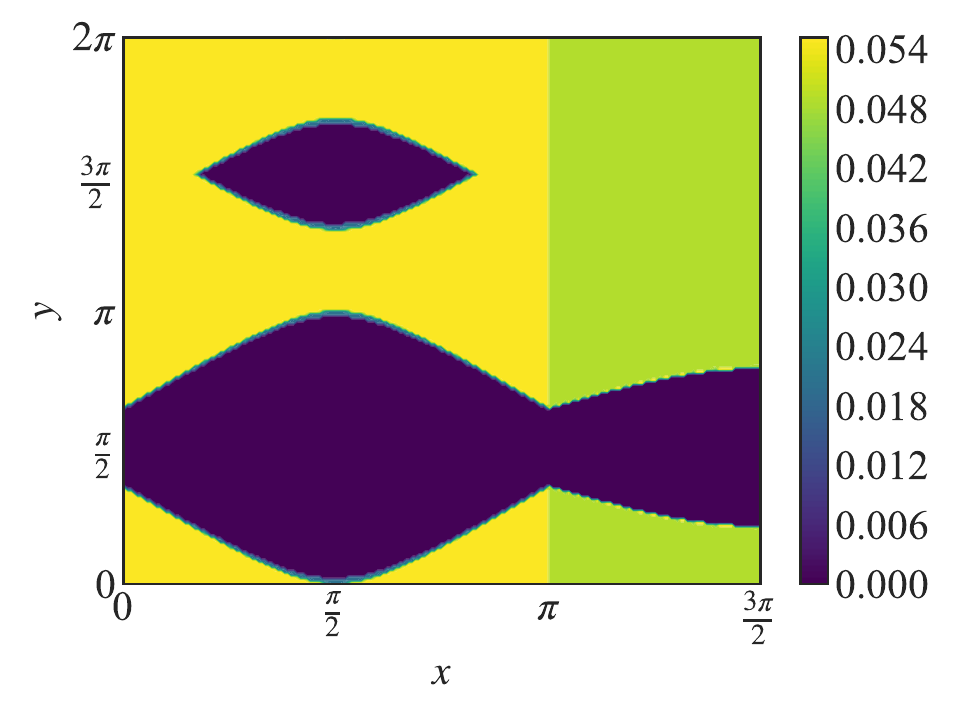}}\quad%
    \caption{Figures of merit in the single cell spin$-(\frac{1}{2}-\frac{1}{2})$ ATIH model, Eq.~\eqref{ham}, under the parameters defined in Eq.\eqref{phase_diagram_1_parameters}. (a) The average Wigner function values, Eq.~\eqref{wigner_arbirary_system}, across the phase space mapping out the complete phase diagram and critical phase boundaries under the equal angle slice approximation. Distinct Wigner function values characterize different phases, varying from positive to negative as influenced by the phase space parameters $(\theta, \varphi)$, demonstrating the utility of the Wigner function in identifying phase transitions within the model. (b) The negativity of the Wigner function, Eq.~\eqref{negativity_wigner}, under the equal angle slice approximation showing its ineffectiveness in this limit in capturing properly the phase diagram~\eqref{phase_diagram_1}. We assess this properly in (c) by adopting a comprehensive mapping in the entire phase space using Monte Carlo~(MC) integration, which confirms the predominance of the negative values of the Wigner function across the phase space, especially significant around the phase boundaries, except in the coherence-free FM and FRI phases. (d) Lower bound concurrence, Eq.~\eqref{LBCC}, splitting the phase diagram in three parts: unentangled region which describes the FM and FRI phases. Maximally entangled region describing the QFO III and IV phases, and an intermediate region that identify the FRU III and FRU IV phases. }
   \label{fig_examples}
\end{figure*}
We begin by examining the phase diagram for the spin$-(\frac{1}{2}-\frac{1}{2})$ ATIH chain, Eq.~\eqref{ham}. In Fig.~\eqref{average_wigner_4_spin_half}, we present the average value of Wigner function, Eq.~\eqref{total_wigner_spin_half}, across the phase space spanned by $(\theta,\varphi)$ under the equal angle slice approximation. The Wigner function can effectively capture the entire phase diagram and all the phase boundaries (c.f Fig.~\eqref{phase_diagram_1}). Furthermore, each phase within the diagram is characterized by distinct values of the Wigner function, which vary from positive to negative based on the parameters $(x,y)$. Here, the equal angle slice approximation is a good simplification for capturing the salient critical features of the spin$-(\frac{1}{2}-\frac{1}{2})$ ATIH model, Eq.~\eqref{ham}, without the need to explore all the phase space which is computationally exhaustive.

Approaching the phase boundaries, the Wigner function tends to maximize either its negative or positive value, depending on the crossing. For instance, the QFO IV - FM crossing is characterized by high-negativity boundary, while the FRI - FRU III boundary is highly positive. This behavior underscores the importance of exploring the Wigner function on both its negative and positive parts to gain a comprehensive understanding of the phase diagram of the system.


The negativity of the Wigner function, Eq.~\eqref{negativity_wigner} is depicted in Fig.~\eqref{negativity_wigner_4_spin_half} under the equal angle slice approximation. The negativity recognize most of the phases with distinctive values, while missing three phase boundaries, i.e. the FRI-$\text{FRU}_{III}$, the $\text{FRU}_{III}-\text{QFO}_{III}$, and the FM-QFO III phase boundaries, which are captured by the positive part of the Wigner function as shown in panel~\eqref{average_wigner_4_spin_half}. Here, the equal angle slice approximation is not adequate to properly quantify the negativity of the Wigner function in the system as we would expect all the phases of matter, except the FM and FRI phases, to have negative Wigner functions. Our expectation comes from the fact that in these phases the Heisenberg nodes responsible for the ``quantum'' effects in the model are described by entangled states~(c.f.~\Cref{eigenstate1_spin_half,eigenstate2_spin_half,eigenstate3_spin_half,eigenstate4_spin_half}), which are known to have a negative Wigner function~\cite{2021_PRR_WigNeg}. We confirm this in panel~\eqref{negativity_wigner_4_spin_half_MC} where we drop the equal angle slice and compute the negativity of the Wigner function, Eq.~\eqref{negativity_wigner}, over the entire phase space spanned by the angles $\{\theta_1,\varphi_1;\theta_2,\varphi_2;\theta_3,\varphi_3;\theta_4,\varphi_4\}$. We see that the Wigner function is negative over the entire phase diagram except in the coherence-free phases, i.e. the FM and FRI phases. Furthermore, the phase boundaries are more pronounced, including the $x\!=\!\pi$ continuous phase transition line, by a maximum amount of negativity.  
\begin{figure*}
    \subfloat[Average of the Wigner function, Eq.~\eqref{wigner_arbirary_system}, under the equal angle slice. \label{eas_spin_one}]{\includegraphics[width=0.23\textwidth]{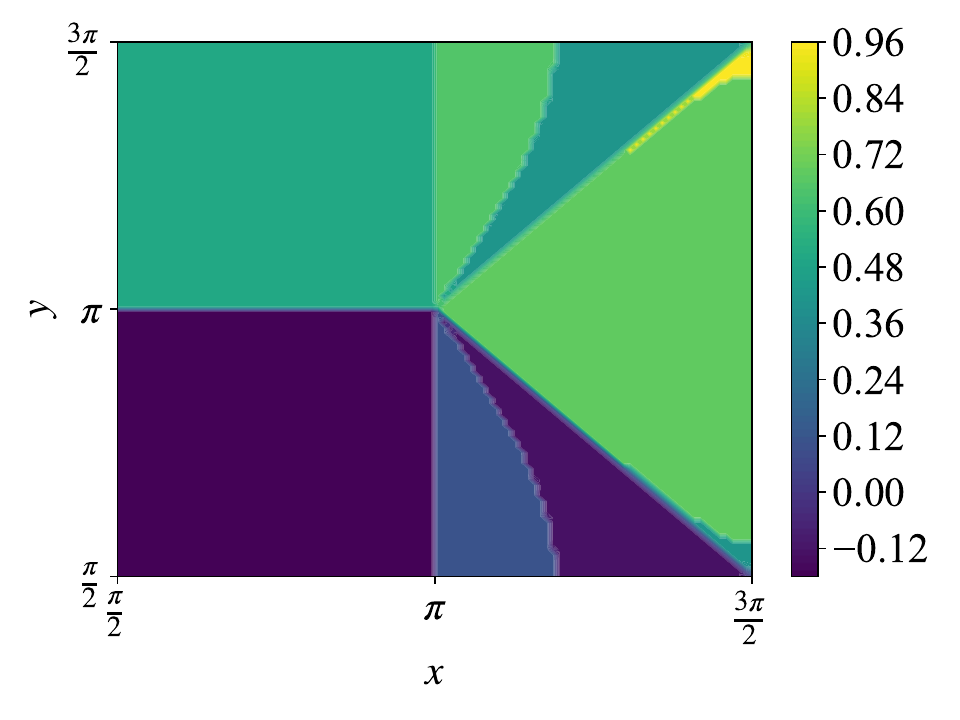}}\quad%
    \subfloat[Average of the Wigner function, Eq.~\eqref{wigner_arbirary_system}, in the entire phase space.\label{mc_spin_one}]{\includegraphics[width=0.23\textwidth]{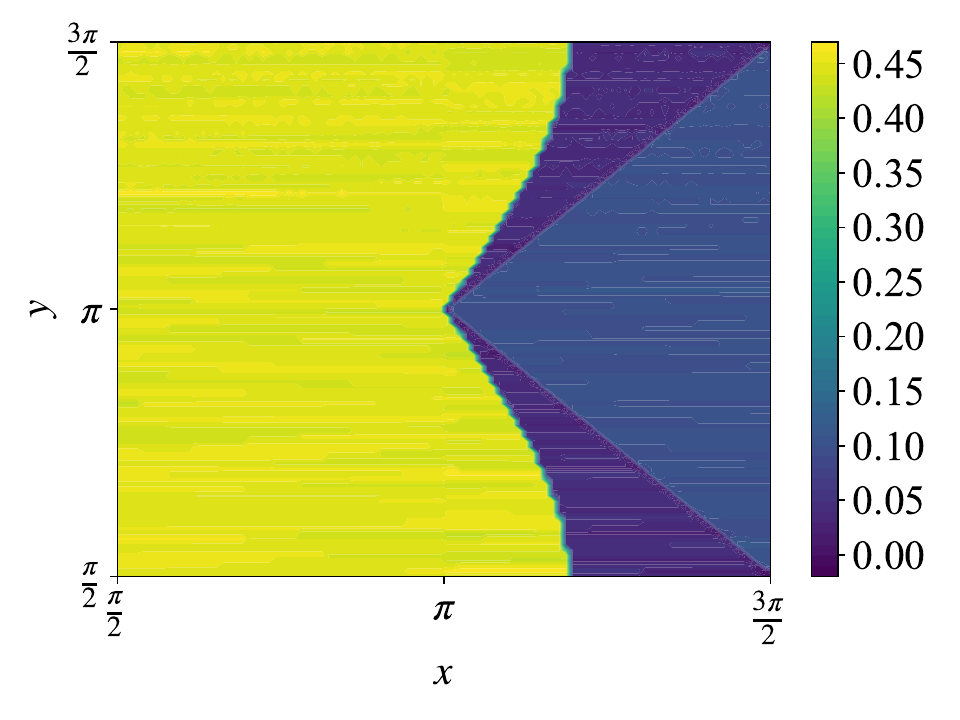}}\quad%
    \subfloat[Negativity of the Wigner function, Eq.~\eqref{negativity_wigner}, using MC integration.\label{negativity_mc_spin_one}]{\includegraphics[width=0.23\textwidth]{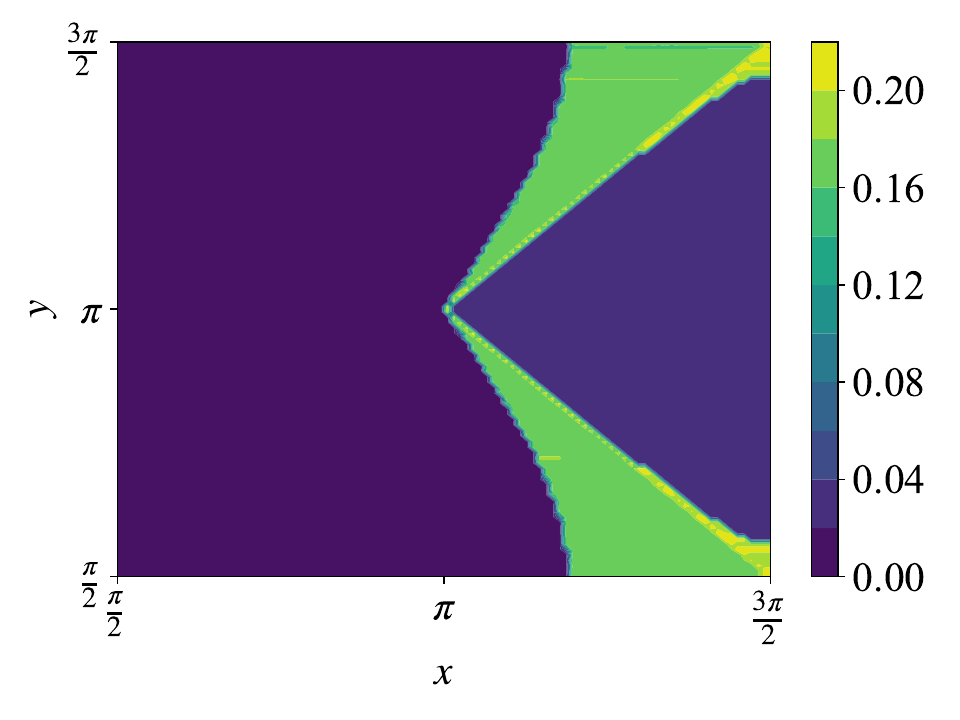}}\quad%
    \subfloat[Lower bound concurrence, Eq.~\eqref{LBCC}.\label{lbc_spin_one}]{\includegraphics[width=0.23\textwidth]{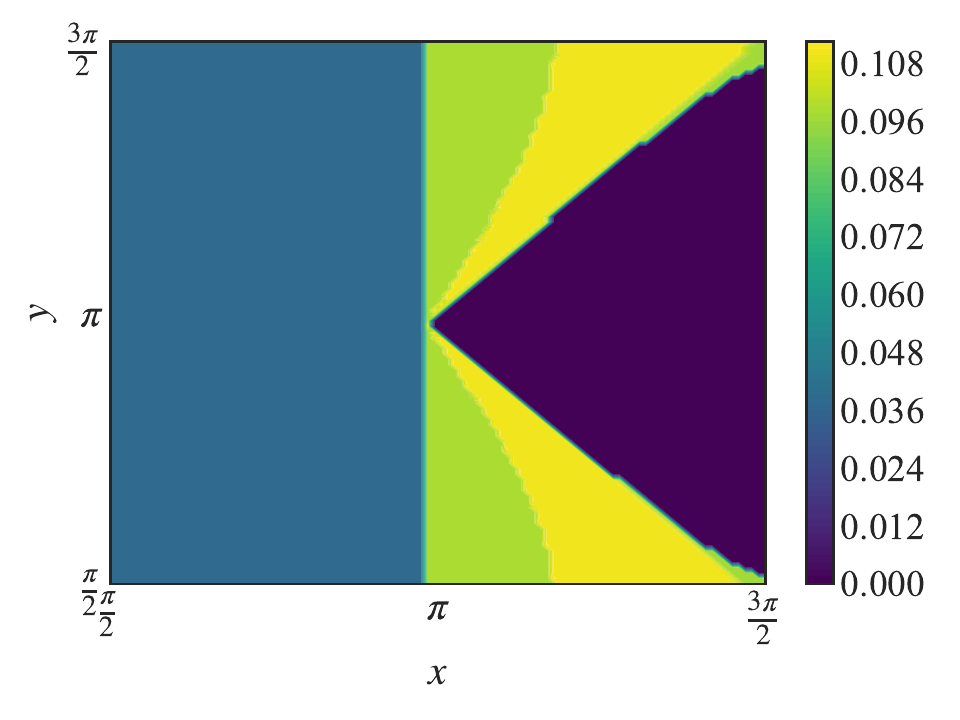}}\quad%
    \caption{Figures of merit in the single cell spin$-(\frac{1}{2}-1)$ ATIH model, Eq.~\eqref{ham}, under the parameters defined in Eq.\eqref{phase_diagram_2_parameters}. (a) Average of the Wigner function, Eq.~\eqref{wigner_arbirary_system}, under the equal slice approximation showing its limitation in revealing the phase diagram~\eqref{phase_diagram_2} for the system, which is due to the bias introduced by the approximation and the information loss in non-uniform regions of phase space. (b) 
    The average Wigner function, Eq.~\eqref{wigner_arbirary_system}, across the entire phase space, providing a clearer description of phase boundaries, particularly between regions of positive and negative Wigner function values. (c+d) Negativity of the Wigner function, Eq.~\eqref{negativity_wigner}, calculated via Monte Carlo integration and the lower bound concurrence, Eq.~\eqref{LBCC}, respectively. Negative values cluster in phases where Heisenberg nodes align with highly entangled Bell states, such as the QFO I and QFO II states. In contrast, the FRU and QFO III phases, associated with generalized-Bell states, show no such negativity despite also being entangled. This suggests that the absence of negativity does not imply the lack of entanglement.}
    \label{fig_full_spin_one}
\end{figure*}
Computing the Wigner function over an 8-dimensional phase space is intractable. Therefore, to generate the data in panel~\eqref{negativity_wigner_4_spin_half_MC} we used statistical approximations via Monte Carlo integration to estimate the values of integrals based on random sampling~\cite{weinzierl2000introduction}. The method is particularly useful for dealing with high-dimensional integrals where traditional numerical integration methods become inefficient or infeasible~\cite{mc_integration_1990}. The method involves generating random points in the domain of the integral and then estimating the integral based on the average value of the function at these points~\cite{newman1999monte}.

The effectiveness of Monte Carlo integration increases with the number of dimensions, making it highly suitable for integrals in spaces with dimensions as high as eight or more. This suitability arises because, unlike many deterministic methods, the convergence rate of Monte Carlo integration does not directly depend on the number of dimensions; it depends only on the number of samples, converging with a rate of $1/\sqrt{N}$, where $N\!=\!2\times10^5$ is the number of random samples. This makes it particularly powerful for tackling complex, high-dimensional problems where the geometry or volume of the integration region is complicated~\cite{kroese2013handbook}.

To assess the versatility of the Wigner function approach, we discuss it in light of the lower bound concurrence, Eq.~\eqref{LBCC}, shown in panel~(\ref{lbc_wigner_4_spin_half}). This entanglement measure characterizes the phase diagram~\eqref{phase_diagram_1} into a entangled and unentangled region. Accordingly, the lower bound concurrence, Eq.~\eqref{LBCC}, splits the phase diagram into three regions: one with zero entanglement, which describes the FM and FRI phases, another with maximum value describing the $\text{QFO}_{III}$ and $\text{QFO}_{IV}$ phases, and an intermediate region describing $\text{FRU}_{III}$ and $\text{FRU}_{IV}$. However, unlike the Wigner function that assign a distinctive value for each phase of matter, the lower bound concurrence, Eq.~\eqref{LBCC}, fails to distinguish between phases of matter in a given entangled (unentangled) region. This shows the utility of the Wigner function approach compared to the lower bound concurrence. 

At a closer inspection of panel~\eqref{negativity_wigner_4_spin_half_MC} and~\eqref{lbc_wigner_4_spin_half}, we see that the negativity of the Wigner function,  Eq.~\eqref{negativity_wigner}, follows a pattern similar to the lower bound concurrence, Eq.~\eqref{LBCC}, but on a different scale. However, while the negativity of the Wigner function remains consistent across all entangled phases, the lower bound concurrence provides a finer distinction, differentiating between maximally entangled QFO states and entangled FRU states. This limitation in the negativity of the Wigner function stems from the Monte Carlo approximation, which computes average values of the integrals and could potentially introduce bias. The reported results support the connection between the entanglement and the presence of negativity in the Wigner function, laying a solid foundation for further exploration of how negativity in phase space correlates with entanglement.
\paragraph*{Inhomogeneous ATIH chain. }
We shift our focus to the inhomogeneous spin$-(\frac{1}{2}\!-\!1)$ chain. Similarly with the homogeneous case, we employ the equal angle slice approximation to explore the phase diagram of the system (c.f.~Fig~\eqref{phase_diagram_2}). Interestingly, this approximation is limited in this case as  shown in Fig~\eqref{eas_spin_one}, where the Wigner function fails to accurately describe the critical properties of the system. The equal angle slice method, like any approximation, sacrifices precision for computational ease. For inhomogeneous quantum systems, this can lead to issues including loss of information in non-uniform areas of the phase space, rendering the results dependent on the chosen slices. Consequently, the approximation does not capture the complete features and detailed structure of the Wigner function.

Given these limitations, we abandon the equal angle slice and instead compute the average Wigner function across the entire phase space spanned by the angles $\{\theta_1,\varphi_1;\theta_2,\varphi_2;\theta_3,\varphi_3;\theta_4,\varphi_4\}$ as illustrated in Fig.~\eqref{mc_spin_one}. This approach provides a more accurate representation of the system's phase diagram. It distinctly highlights the phase boundaries, particularly those delineating regions of positive and negative Wigner function, such as between QFO II - QFO III. However, the $x=\pi$ phase boundary is less visible due to its positioning between the FRU and QFO phases, which are characterized by a predominantly high positive Wigner function, complicating its detection.

Fig.~\eqref{negativity_mc_spin_one} and~\eqref{lbc_spin_one} highlights the negative regions of the Wigner function across the entire phase space using the Monte Carlo approximation for computing integrals and the lower bound concurrence, respectively, for the spin$-(\frac{1}{2}-1)$ ATIH model. Negative values are localized in phases of matter where the Heisenberg nodes correspond to Bell states, specifically the QFO I and QFO II states (see~\Cref{qfo1_spin1,qfo2_spin1}), which exhibit high entanglement as shown in panel~\eqref{lbc_spin_one}. However, the negativity of the Wigner function fail to distinguish between the FRU and QFO III phases, which involve generalized-Bell states (see~\Cref{fru_spin1,qfo3_spin1}) and are also entangled according to panel~\eqref{lbc_spin_one}. This suggests that the lack of negative regions in the Wigner function does not necessarily indicate an absence of entanglement in those areas of phase space. These results call for a profound analysis of Wigner negativity and entanglement in high-dimensional states. The role of Wigner negativity in quantum systems remains a vibrant open question, particularly in exploring its potential to measure entanglement and other quantum informational resources. The measurement and practical utilization of Wigner negativity can have implications for quantum computing and quantum communication protocols.
 \section{Conclusion }
We have analyzed the phase diagrams of spin$-(\frac{1}{2}-\frac{1}{2})$ and spin$-(\frac{1}{2}-1)$ Ising-Heisenberg chains through the lens of the Wigner function. Our results have illuminated the distinct roles played by the negative and positive parts of the Wigner function in revealing the complex phase structures and boundaries of these chains. The comparative analysis with entanglement concurrence has not only validated our findings but also highlighted the strengths and limitations of each approach in revealing the phase diagram of the systems.

The versatility of the equal angle slice approximation in the homogeneous spin$-(\frac{1}{2}-\frac{1}{2})$ chain, as opposed to the necessity of a full phase space integration for the inhomogeneous spin$-(\frac{1}{2}-1)$ chain, underlines the crucial influence of the quantum system homogeneity on phase space analysis. These insights pave the way for more nuanced approaches to studying quantum systems, particularly in understanding the interplay between system properties and the choice of phase space methods.

Our results lay the foundation for the inspection of a profound relationship between the negativity of the Wigner function and entanglement in quantum many-body systems. This investigation aims to unravel the underlying properties of quantum matter, with a specific focus on how entanglement arise in various quantum phases and its manifestation in phase space. By bridging the gap between Wigner function negativity and quantum entanglement, we aspire to develop a more comprehensive framework for detecting and characterizing the properties of quantum matter. This endeavor will not only enrich our understanding of quantum phase transitions but also contribute significantly to the broader field of quantum physics, potentially influencing the development of quantum computing and information processing technologies.\\
\section*{Acknowledgements}
Z.M. acknowledge support from the National Science Center (NCN), Poland, under Project No.$~2020/38/E/ST3/00269$. S.D. acknowledges support from the John Templeton Foundation under Grant No. 62422. B.G. acknowledge support from the National Science Center (NCN), Poland, under Project No.$~2022/47/B/ST6/02380$. This research is supported through computational resources of \href{https://hpc.marwan.ma/index.php/en/}{HPC-MARWAN} provided by CNRST. 
\appendix
\section{\label{spect_ATIH}Spectrum of the ATIH chain}
We report here the ground state properties of the ATIH model for the homogeneous $S=\frac{1}{2}$ and inhomogeneous $S=1$.
\paragraph*{\bf Homogeneous ATIH chain.}
To study the phase diagram of the ATIH chain we diagonalize the Hamiltonian, Eq.~\eqref{ham}. The introduction of the notations $J_+ = J_x + J_y$ and $J_- = J_x - J_y$ enables us to recast the Hamiltonian as
\begin{equation}
    H_{i,i+1}=\text{diag}(\epsilon_{+}^{1},\epsilon_{+}^{2},\epsilon_{-}^{2},\epsilon_{-}^{1}),
\end{equation}
where \text{diag}() represent the diagonal elements of the Hamiltonian, Eq.~\eqref{ham}, with the eigenvalues:
\begin{align}
\epsilon_{\pm}^{1}( \sigma_{i} , \sigma_{i+1}) &=  \gamma + \frac{J_z}{4} \pm \frac{1}{4} \sqrt{16\alpha^2  + J_{-}^2 }, \label{e1} \\
\epsilon_{\pm}^{2}( \sigma_{i} , \sigma_{i+1}) &= \gamma  -\frac{J_z}{4} \pm \frac{1}{4}J_{+}, \label{e2}
\end{align}
with 
\begin{align} 
\gamma & \equiv  \gamma( \sigma_{i}, \sigma_{i+1}) = J\sigma_{i}  \sigma_{i+1}+\frac{h_{0}}{2}( \sigma_{i} +  \sigma_{i+1}), \\
\alpha & \equiv \alpha( \sigma_{i} , \sigma_{i+1}) = J( \sigma_{i} + \sigma_{i+1})  + h.
 \end{align}
The corresponding eigenstates in terms of standard basis ${|++\rangle , |+-\rangle , |-+\rangle , |--\rangle}$ are given respectively by: 
 \begin{align}
\ket{\phi_1}&= \frac{1}{ \sqrt{ 1  + e_{1}^2}}(e_{1}\ket{++}+\ket{--}), \label{eigenstate1_spin_half}\\
\ket{\phi_2}&= \frac{1}{ \sqrt{2}}(\ket{+-}+\ket{-+}), \label{eigenstate2_spin_half}\\
\ket{\phi_3}&= \frac{1}{ \sqrt{2}}(-\ket{+-}  + \ket{-+}), \label{eigenstate3_spin_half}\\
\ket{\phi_4}&= \frac{1}{ \sqrt{ 1  + e_{2}^2}}(e_{2}\ket{++} + \ket{--}), \label{eigenstate4_spin_half}
\end{align}
with
\begin{align}
    e_{1} & \equiv e_{1}( \sigma_{i}, \sigma_{i+1}) =  \frac{\sqrt{16\alpha^2  + J_{-}^2} + 4\alpha}{J_{-}}, \\
    e_{2} & \equiv e_{2}( \sigma_{i}, \sigma_{i+1}) = \frac{-\sqrt{16\alpha^2  + J_{-}^2} + 4\alpha}{J_{-}}.
    \label{e1}
\end{align}
The phase diagram of ATIH spin-$(\frac{1}{2}-\frac{1}{2})$ displays a rich variety of quantum states and phase transitions due to the complex interplay of the Ising and Heisenberg interactions. Four distinct quantum phases of matter are identified: Quantum Ferromagnetic (QFO), Frustrated Ferromagnetic (FRU), Ferromagnetic (FM), and Ferrimagnetic (FRI). Their corresponding quantum state are given by
\begin{align}
   |\text{QFO}_{I}\rangle &=  \prod_{k=1}^N  |+ , \phi_{1}(+,+)\rangle_{k},\label{qfo1}\\
   |\text{QFO}_{II}\rangle &=  \prod_{k=1}^N  |+ , \phi_{4}(+,+)\rangle_{k},\label{qfo2}\\
   |\text{QFO}_{III}\rangle &=  \prod_{k=1}^N  |+ , \phi_{2}\rangle_{k},\label{qfo3}\\
   |\text{QFO}_{IV}\rangle &=  \prod_{k=1}^N  |+ , \phi_{3}\rangle_{k},\label{qfo4}\\
   |\text{FRU}_{I}\rangle &=  \prod_{k=1}^{N/2}  |+ , \phi_{1}(+,-),-,\phi_{1}(+,-)\rangle_{k},\label{fru1}\\
   |\text{FRU}_{II}\rangle &=  \prod_{k=1}^{N/2}  |+ , \phi_{1}(+,-),-,\phi_{1}(+,-)\rangle_{k},\label{fru2}\\
   |\text{FRU}_{III}\rangle &=  \prod_{k=1}^{N/2}  |+ , \phi_{4},-,\phi_{4}\rangle_{k},\label{fru3}\\
   |\text{FRU}_{IV}\rangle &=  \prod_{k=1}^{N/2} |+ , \phi_{3},-,\phi_{3}\rangle_{k},\label{fru4}
\end{align}
where the product is carried over all sites. The first element in the product corresponds to the Ising edge which can take only two possible values $\pm 1$, while the second element represents the Heisenberg edge.
The Quantum Ferromagnetic phases are categorized into four types (QFO I-IV), each characterized by unique ground state vector products over all lattice sites, and they differ primarily in the orientation probabilities of spins in the Heisenberg edges, which are determined by the functions $e_1$ and $e_2$, Eq.~\eqref{e1}. These functions highlight the quantum nature of the ferromagnetic state by indicating that up and down orientations have distinct probabilities. Meanwhile, the Frustrated Ferromagnetic states (FRU I-IV) demonstrate non-degenerate characteristics with spin frustration present, indicative of the competition between different interactions within the lattice.

The FM and FRI states are defined by the magnetizations of the Ising and Heisenberg edges. In the FM state, spins are aligned in the same direction, leading to a uniform magnetic moment, whereas in the FRI state, there is an alternating pattern of spins which results in a net magnetization that is less than the saturation magnetization. These phases emerge from the interactions that promote parallel and anti-parallel spin alignments, respectively, illustrating the diverse magnetic behaviors that can arise from the interplay of Ising and Heisenberg interactions in low-dimensional quantum systems.
\paragraph*{\bf Inhomogeneous ATIH chain. }
We consider the Heisenberg edge with $S=1$ in the ATIH model, Eq.~\eqref{ham}. Accordingly, the diagonalized Hamiltonian can be written as
\begin{equation}
     H_{i,i+1}=\text{diag}(\epsilon_{+}^1, \epsilon_{+}^2, \epsilon_{+}^3, \epsilon_{+}^4, \epsilon^5, \epsilon_{-}^4, \epsilon_{-}^3, \epsilon_{-}^2, \epsilon_{-}^1),
\end{equation}
with the following eigenvalues
\begin{align}
\epsilon_{\pm}^1( \sigma_{i} , \sigma_{i+1}) &= \pm2\alpha + \gamma + J_{z}, \label{e19} \\
\epsilon_{\pm}^2( \sigma_{i} , \sigma_{i+1}) &= \pm\alpha + \gamma + \frac{1}{2}J_{+}, \label{e28}\\
\epsilon_{\pm}^3( \sigma_{i} , \sigma_{i+1}) &= \pm\alpha + \gamma - \frac{1}{2}J_{+}, \label{e37} \\
\epsilon_{\pm}^4( \sigma_{i} , \sigma_{i+1}) &=  \gamma - \frac{1}{2}J_{z} \pm\frac{1}{2} \sqrt{(J_{z}^2  +  J_{+}^2)}, \label{e46}\\
\epsilon^{5}( \sigma_{i} , \sigma_{i+1}) &=  \gamma -J_{z}. \label{e5}
\end{align}
The corresponding eigenstates in the computational basis  $\ket{-1,-1}$, $\ket{-1,0}$, $\ket{-1,1}$, $\ket{0,-1}$, $\ket{0,0}$, $\ket{0,1}$, $\ket{1,-1}$, $\ket{1,0}$, $\ket{1,1}$ are given by:
\begin{align}
|\varphi_{+}^1\rangle &= |1,1\rangle, \\
|\varphi_{-}^1\rangle &= |-1,-1\rangle, \\
|\varphi_{+}^2\rangle &= \frac{1}{\sqrt{2}}(|1,0\rangle  +  |0,1\rangle), \\
|\varphi_{-}^2\rangle &= \frac{1}{\sqrt{2}}(|0,-1\rangle  +  |-1,0\rangle), \\
|\varphi_{+}^3\rangle &= \frac{1}{\sqrt{2}}(-|1,0\rangle  +  |0,1\rangle), \\
|\varphi_{-}^3\rangle &= \frac{1}{\sqrt{2}}(-|0,-1\rangle  +  |-1,0\rangle),\\
|\varphi_{+}^4\rangle &= \frac{1}{\sqrt{2 + f_{1}^2}}(|1,-1\rangle  +  |-1,1\rangle 
 + f_{1}|0,0\rangle), \\
|\varphi_{-}^4\rangle &= \frac{1}{\sqrt{2 + f_{2}^2}}(|1,-1\rangle  +  |-1,1\rangle 
 + f_{2}|0,0\rangle), \\
 |\varphi^5\rangle &= \frac{1}{\sqrt{2}}(-|1,-1\rangle  +  |-1,1\rangle),
\label{eig2s1half}
\end{align}
where 
\begin{align}
f_{1} & = \frac{J_z + \sqrt{J_z^2  + 8J_x^2}}{2J_x}, &
f_{2} & = \frac{J_z - \sqrt{J_z^2  + 8J_x^2}}{2J_x}.
\end{align}
Accordingly, the phase diagram of the ATIH spin-($\frac{1}{2}-1$) chain, Eq.~\eqref{ham}, is characterized by a series of distinct phases: Ferromagnetic (FM), Ferrimagnetic (FRI), Quantum Ferromagnetic (QFO), and Frustrated (FRU) states, as well as Quantum Ferrimagnetic (QFI) states. described as follows
\begin{align}
|\text{FM}\rangle &=  \prod_{k=1}^{N} |+ , \varphi_{+}^1\rangle_{k},\label{fm_spin1} \\
|\text{FRI}\rangle &=  \prod_{k=1}^{N} |- , \varphi_{+}^1\rangle_{k},\label{fri_spin1} \\
|\text{QFO}_{I}\rangle &=  \prod_{k=1}^N  |+ , \varphi_{+}^2\rangle_{k},\label{qfo1_spin1}\\
|\text{QFO}_{II}\rangle &=  \prod_{k=1}^N  |+ , \varphi_{+}^3\rangle_{k},\label{qfo2_spin1}\\
|\text{QFO}_{III}\rangle &=  \prod_{k=1}^N  |+ , \varphi_{-}^4\rangle_{k},\label{qfo3_spin1}\\
|\text{QFI}_{I}\rangle &=  \prod_{k=1}^N  |- , \varphi_{+}^2\rangle_{k},\label{qfi1_spin1}\\
|\text{QFI}_{II}\rangle &=  \prod_{k=1}^N  |- , \varphi_{+}^3\rangle_{k},\label{qfi2_spin1}\\
|\text{FRU}\rangle &=  \prod_{k=1}^{N/2}  |+ , \varphi_{-}^4 , -\varphi_{-}^4\rangle_{k}, \label{fru_spin1}
\end{align}
where the product is carried over all the sites. The first element in the product corresponds to the Ising edge which can take only two possible values $\pm 1$, while the second element represents the Heisenberg edge. 

The FM state exhibits uniform spin alignment, resulting in maximal magnetization across the chain. Conversely, the FRI state has alternating spin alignment between the Ising spins and the Heisenberg chains, leading to a reduced net magnetization compared to the FM state. The Quantum Ferromagnetic phases are divided into three types (QFO I-III) and are distinguished by their spin configurations and magnetization behaviors, which are functions of the Heisenberg edge's interaction parameters. These QFO states showcase the quantum effects inherent in the Heisenberg interaction, where spins are entangled, yielding non-classical magnetic properties. 

Quantum Ferrimagnetic states (QFI I and II) emerge when the system displays magnetic order that is intermediate between the FM and FRI states, characterized by non-integral values of the total magnetization per unit cell. This indicates a tough balance between ferromagnetic and antiferromagnetic interactions. The FRU state represents a magnetically disordered phase where the spins are subject to geometric frustration, preventing them from settling into a regular pattern. This state is particularly notable in systems with competing interactions where the geometric arrangement of the spins precludes simultaneous minimization of all interaction energies, leading to a degeneracy of ground states and a lack of long-range magnetic order.
\section{\label{ATIH_thermo}The ATIH chain thermodynamics}
\begin{figure*}
    \subfloat[Single site magnetization $\langle \sigma_i \rangle$]{\includegraphics[width=0.33\textwidth]{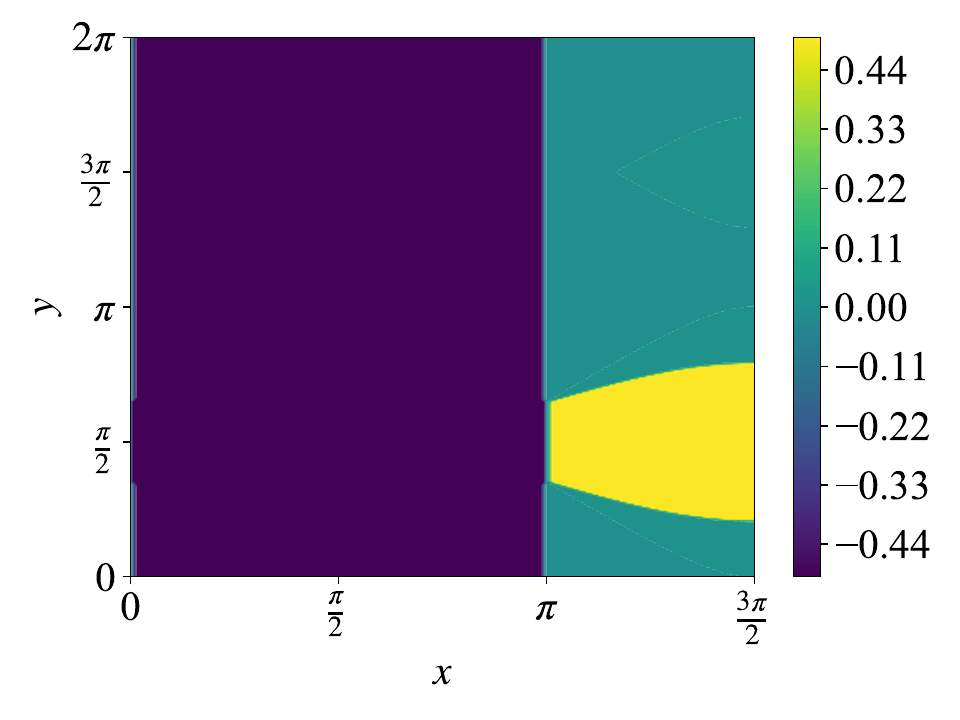}}%
    \subfloat[Two-site Ising correlation function $\langle \sigma_i^z \sigma_{i+1}^z \rangle$]{\includegraphics[width=0.33\textwidth]{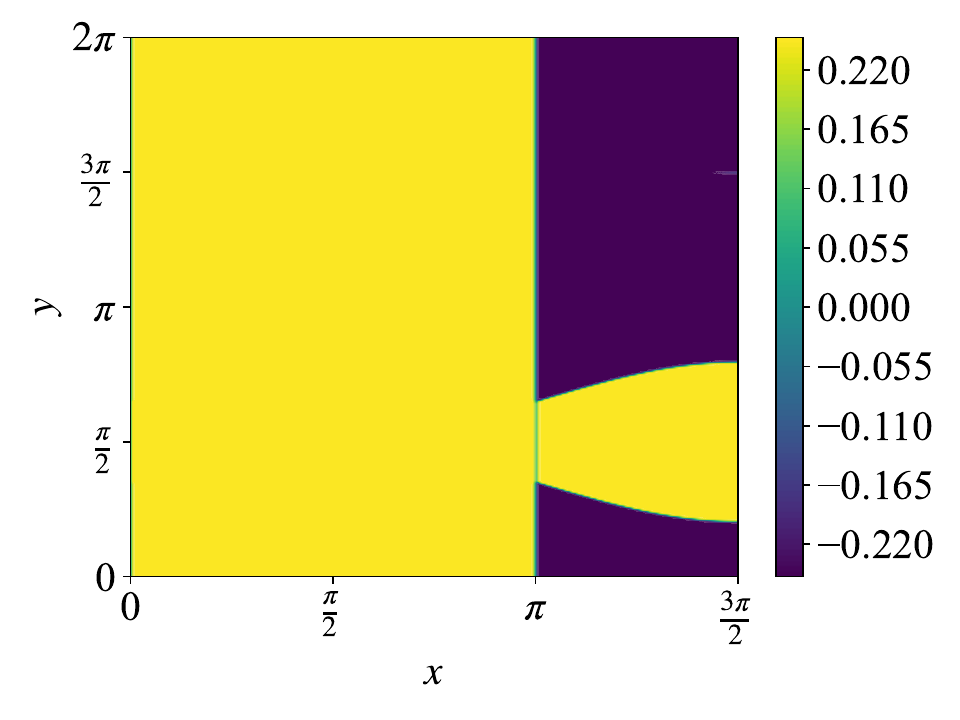}}%
    \subfloat[Single site magnetization $\langle S^z_a \rangle$]{\includegraphics[width=0.33\textwidth]{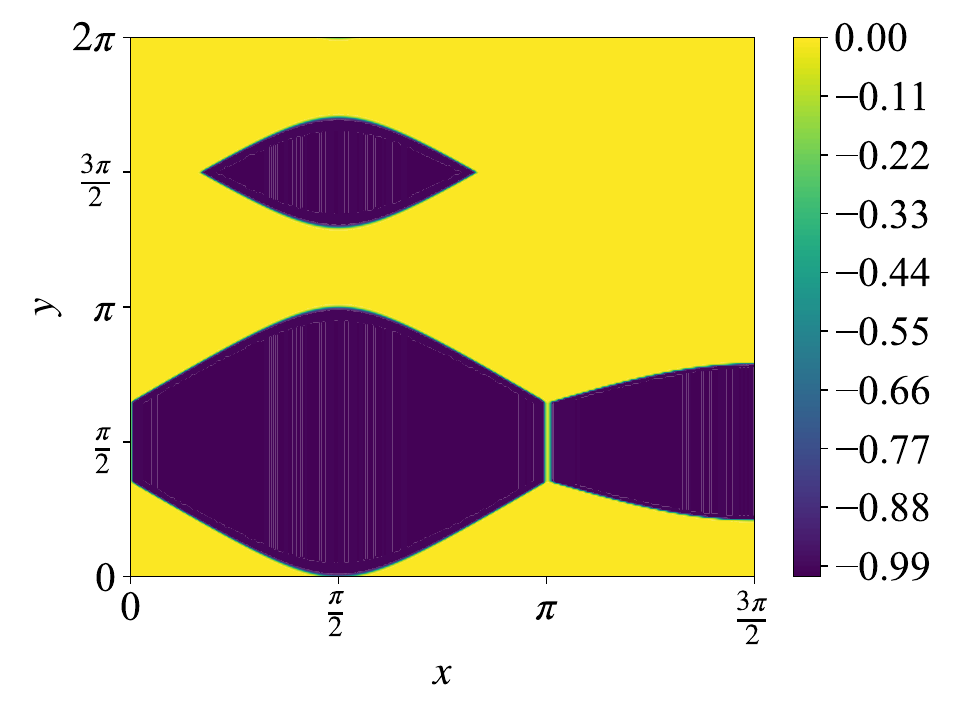}}\\
    \subfloat[Two-site Heisenberg correlation function $\langle S_a^x S_{b}^x \rangle$]{\includegraphics[width=0.33\textwidth]{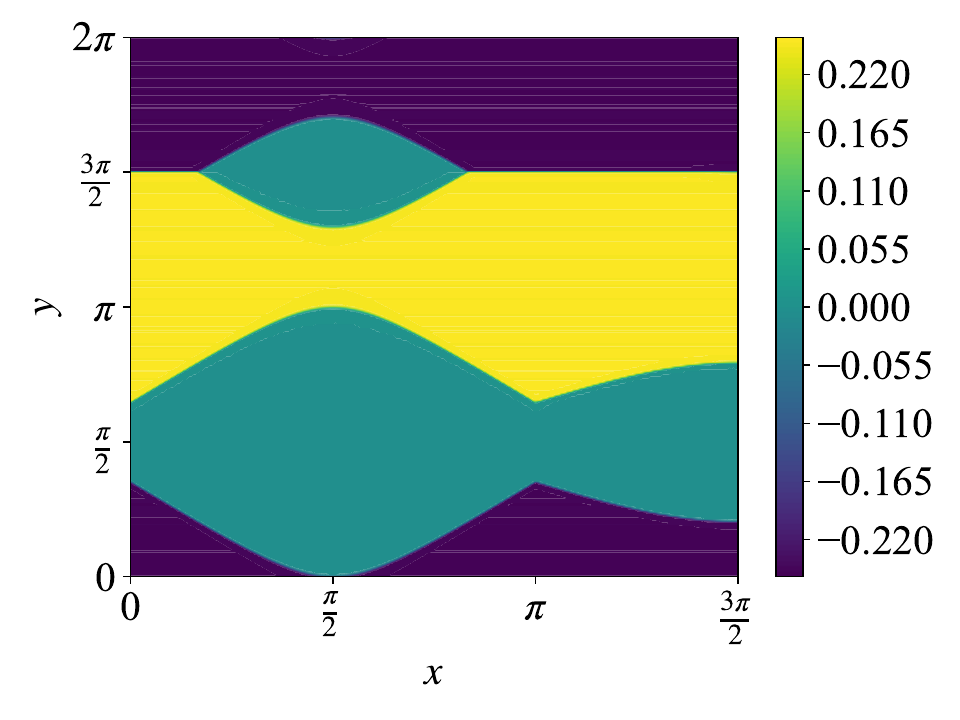}}%
    \subfloat[Two-site Heisenberg correlation function  $\langle S_a^z S_{b}^z \rangle$]{\includegraphics[width=0.33\textwidth]{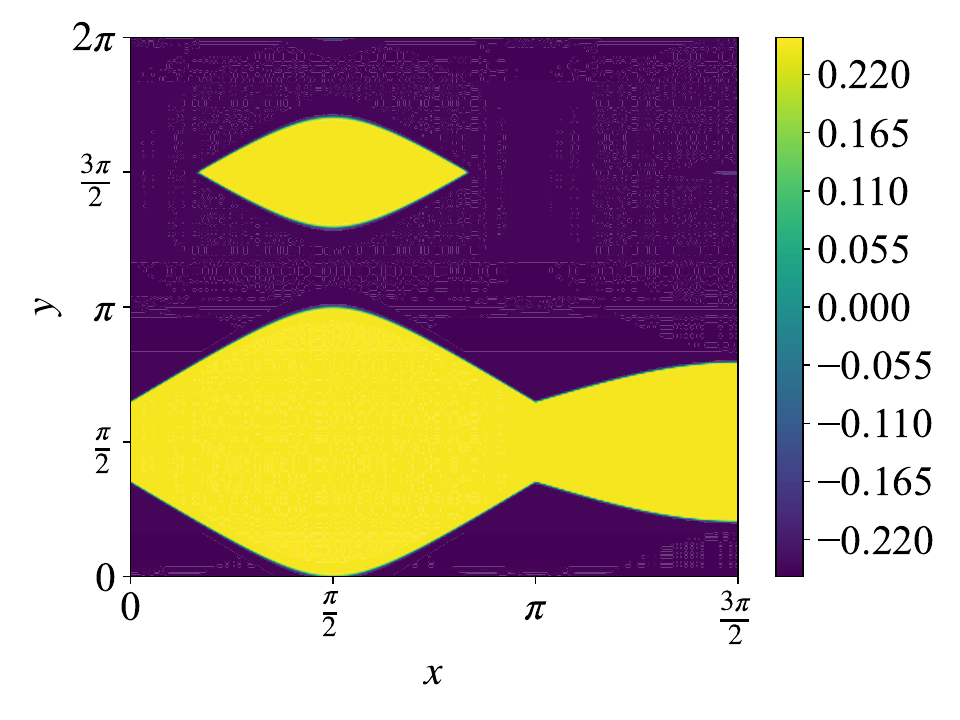}}%
    \subfloat[Two-site Heisenberg correlation function  $\langle S_a^z \sigma_{i}^z \rangle$]{\includegraphics[width=0.33\textwidth]{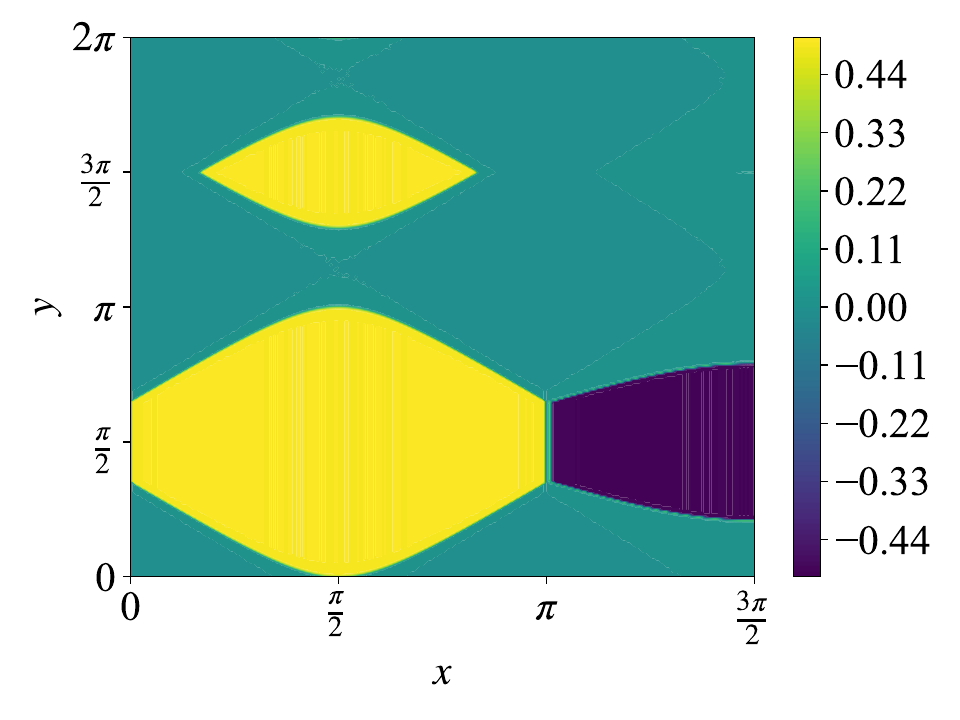}}
    \caption{Correlation function of the ATIH chain in the spin$-(\frac{1}{2}-\frac{1}{2})$ case }
    \label{corr_func_spin_half}
\end{figure*}
\begin{figure*}
    \subfloat[Single site magnetization $\langle \sigma_i \rangle$]{\includegraphics[width=0.33\textwidth]{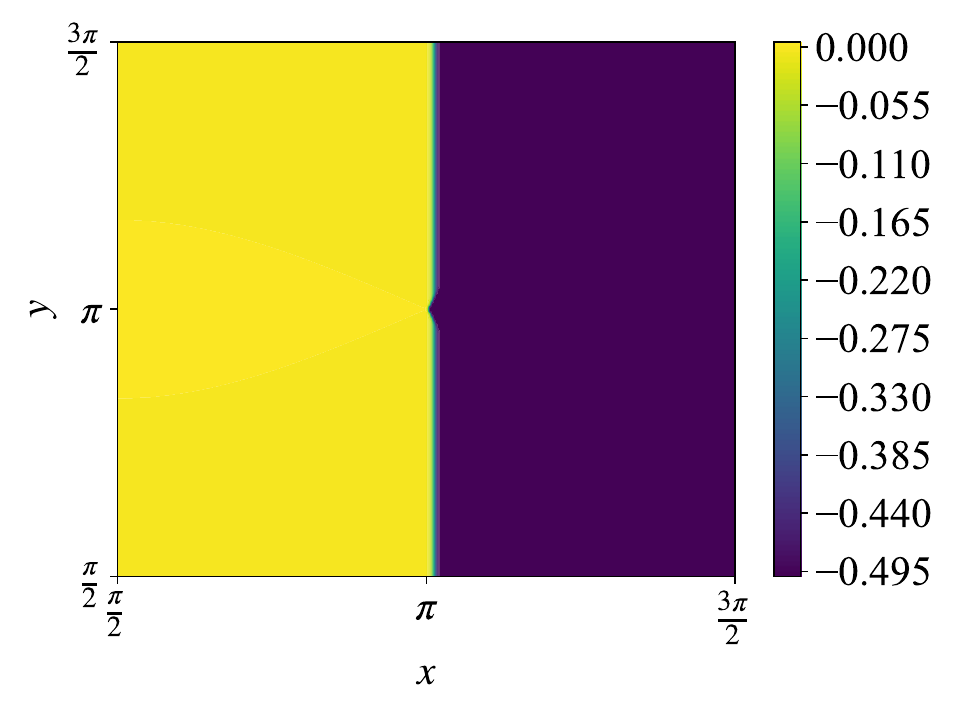}}%
    \subfloat[Two-site Ising correlation function $\langle \sigma_i^z \sigma_{i+1}^z \rangle$]{\includegraphics[width=0.33\textwidth]{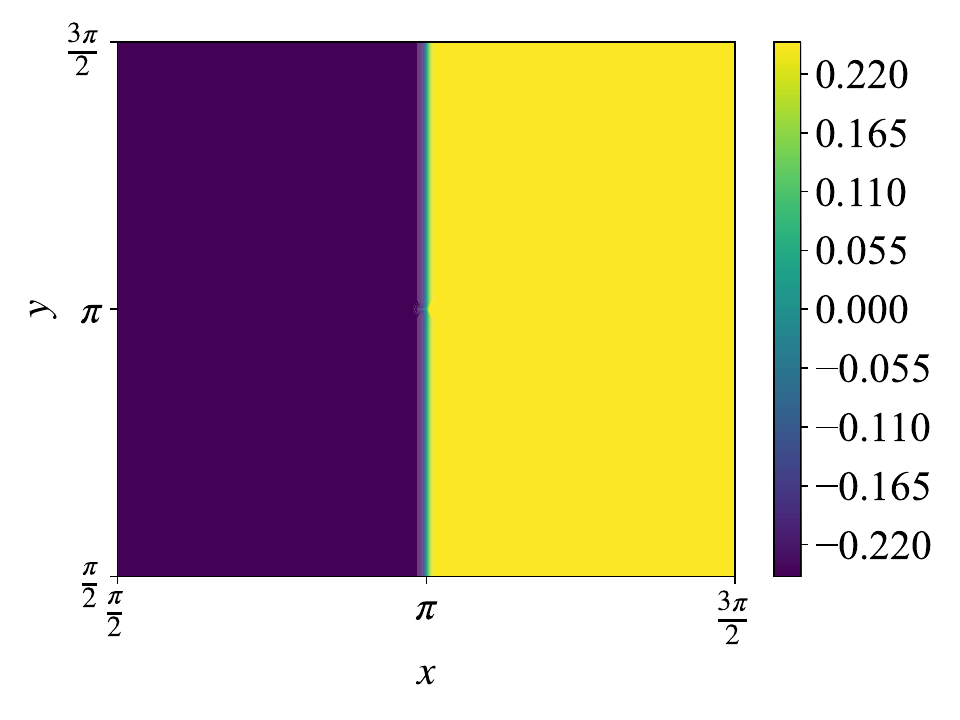}}%
    \subfloat[Single site magnetization $\langle S^z_a \rangle$]{\includegraphics[width=0.33\textwidth]{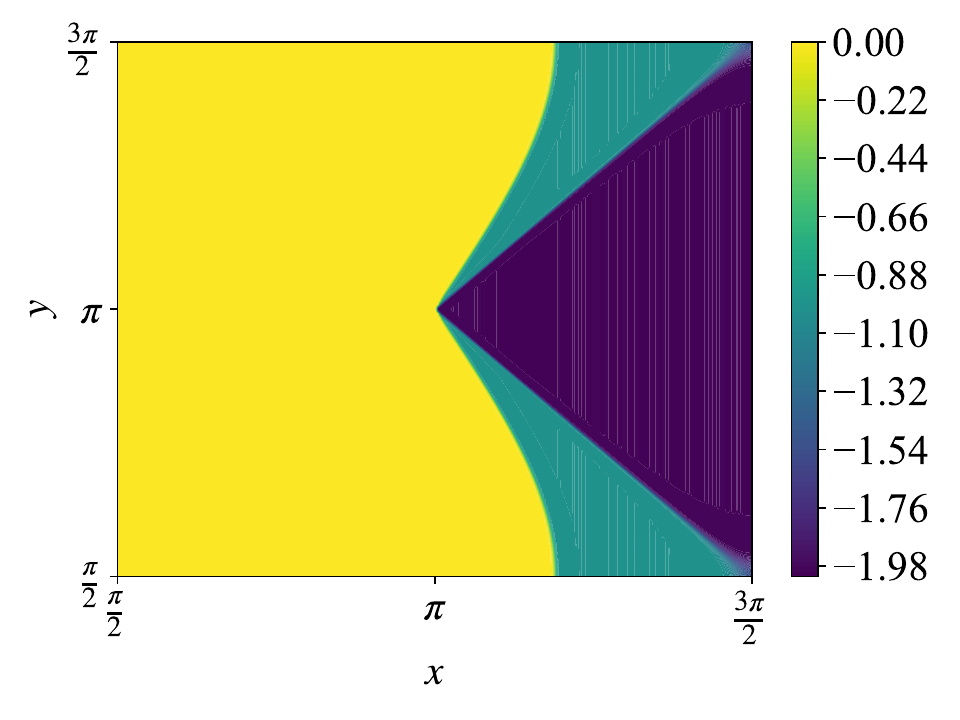}}\\
    \subfloat[Two-site Heisenberg correlation function $\langle S_a^x S_{b}^x \rangle$]{\includegraphics[width=0.33\textwidth]{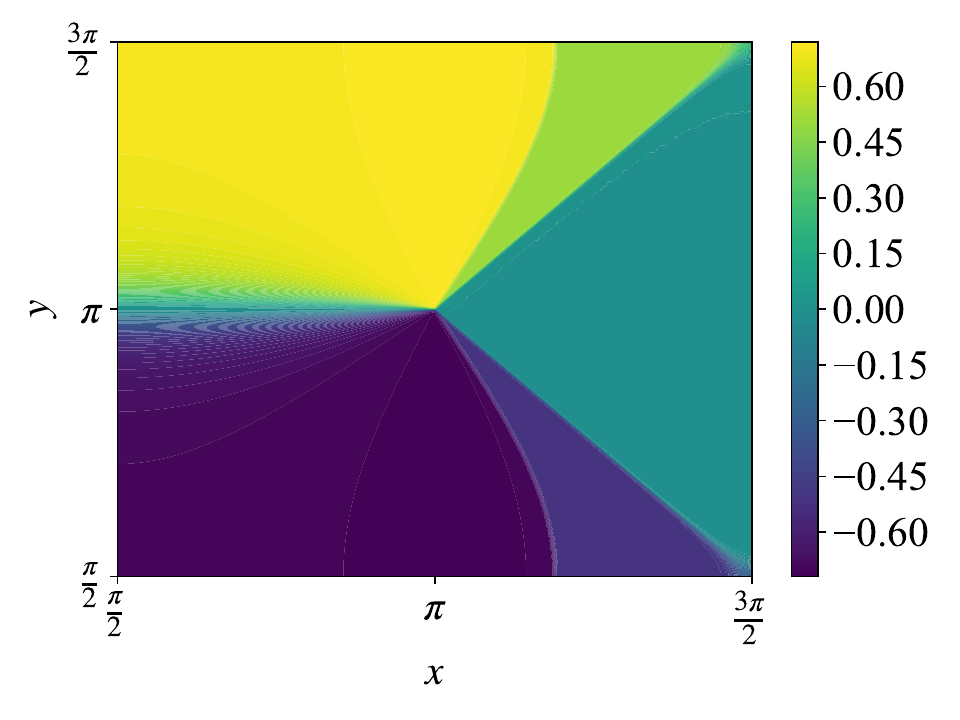}}%
    \subfloat[Two-site Heisenberg correlation function  $\langle S_a^z S_{b}^z \rangle$]{\includegraphics[width=0.33\textwidth]{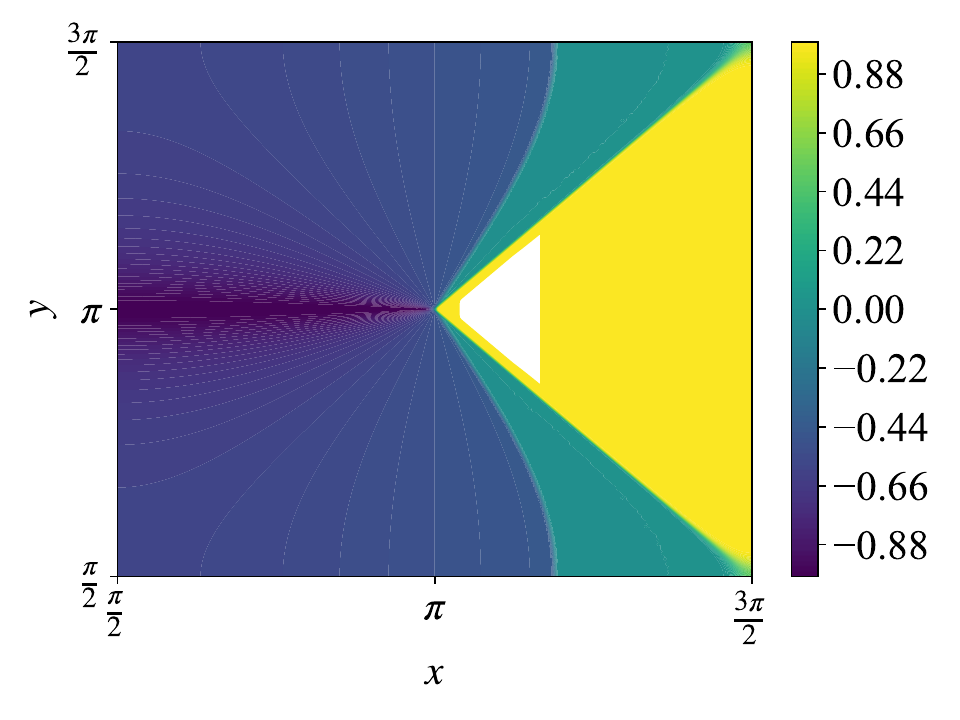}}%
    \subfloat[Two-site Heisenberg correlation function  $\langle S_a^z \sigma_{i}^z \rangle$]{\includegraphics[width=0.33\textwidth]{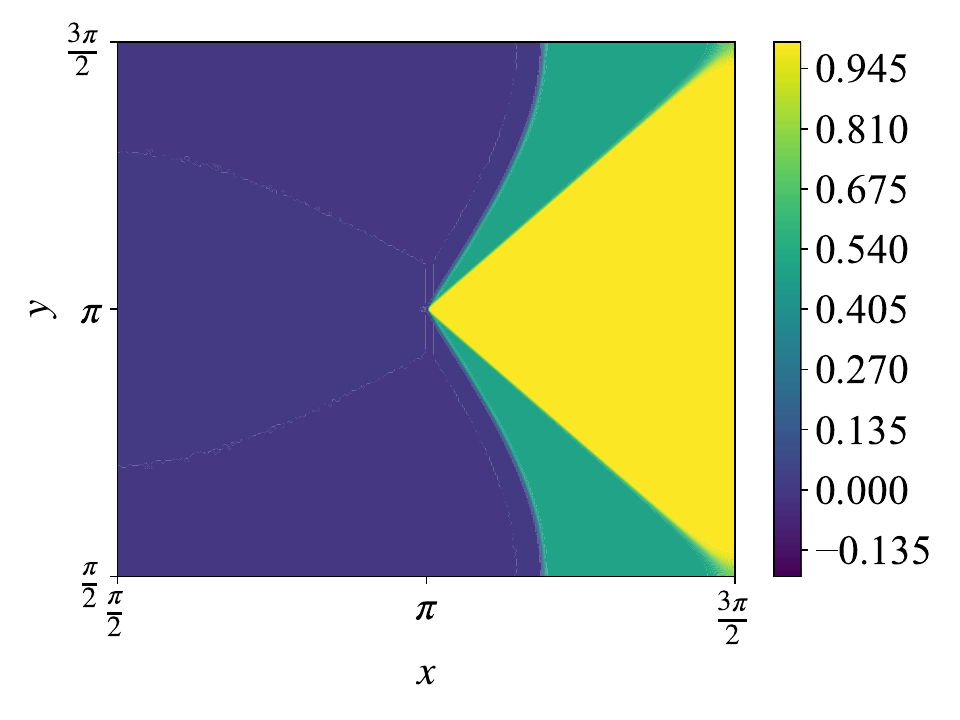}}
    \caption{Correlation function of the ATIH chain in the spin$-(\frac{1}{2}-1)$ case }
    \label{corr_func_spin_one}
\end{figure*}

The thermodynamic properties of the ATIH chain are examined using the decoration-iteration transformation, which is a method initially developed by Fisher~\cite{fisher1959transformations} allows the complex ATIH chain to be mapped into an effective Ising chain~(c.f.~Fig.~\eqref{fig_lattice}) described by the Hamiltonian
\begin{equation}
    \mathcal{H}_{eff} = J_{eff} \sigma_{i}\sigma_{i+1} + h_{eff}(\sigma_{i} + \sigma_{i+1}),
\end{equation}
thereby enabling a simplification of the partition function which can be written as function of the Boltzmann weights  $\tilde{w}(\sigma_i, \sigma_{i+1})$  as
\begin{equation}
   \mathcal{Z} = \sum_{\{\sigma_i\}} \prod_{i=1}^{N}\tilde{w}(\sigma_i, \sigma_{i+1}),
\end{equation}
where $\tilde{w}(\sigma_{i}, \sigma_{i+1})=f_{eff} e^{-\beta \mathcal{H}_{eff} }$ essentially captures the interactions between the Ising spins and the decorated Heisenberg spins, as well as the external magnetic field's effects on the system. The parameters $ f_{eff}, J_{eff} $ and $h_{eff}$ are given by:
\begin{align}
f_{eff}^2 &=w\left(\frac{1}{2}, \frac{-1}{2}\right) \sqrt{w\left(\frac{1}{2}, \frac{1}{2}\right)w\left(\frac{-1}{2}, \frac{-1}{2}\right)}, \\
J_{eff} &= \frac{4}{\beta}\ln\left[\frac{w\left(\frac{1}{2}, \frac{1}{2}\right)w\left(\frac{-1}{2}, -\frac{1}{2}\right)}{\left(w\left(\frac{1}{2}, -\frac{1}{2}\right)\right)^2}\right], \\
 h_{eff} &= \frac{1}{2 \beta}\ln\left[\frac{w\left(\frac{1}{2}, \frac{1}{2}\right)}{w\left(\frac{-1}{2}, \frac{-1}{2}\right)}\right].
\end{align}
The explicit formula for $w(\sigma_{i}, \sigma_{i+1})$  is:
\begin{equation}
w(\sigma_{i}, \sigma_{i+1}) =
\begin{cases}
2e^{-\beta(\gamma + \frac{1}{4} J_z)} \cosh \left(\frac{\beta}{4} \sqrt{16\alpha^2 + J^2}\right)\\
\quad + 2e^{-\beta(\gamma - \frac{1}{4} J_z)} \cosh\left(\frac{\beta}{4} J^+\right), \quad \textbf{\text{for}} \quad   S = \frac{1}{2}
.\\

e^{-\beta\gamma} \left(e^{\beta J_z} + 2e^{\frac{\beta}{2} J_z} \cosh\left(\frac{\beta}{2} \sqrt{J_z^2 
  + 2J_+^2}\right) \right)\\
\quad + 4\cosh\left(\frac{\beta}{2} J_{+}\right)\cosh(\beta\alpha)\\
\quad + 2e^{-\beta J_z} \cosh(2\alpha),

\quad \textbf{\text{for}} \quad   S=1.
\end{cases}
\end{equation}
The correlation functions at the Ising and Heisenberg sites follow immediately~\cite{Lee_1973}. The magnetization at the Ising site can be written as
\begin{equation}
    \langle \sigma^z \rangle = \frac{1}{2} \frac{w(\frac{1}{2},\frac{1}{2})-w(-\frac{1}{2},-\frac{1}{2})}{|w(\frac{1}{2},\frac{1}{2})-w(-\frac{1}{2},-\frac{1}{2})|} \frac{1}{\sqrt{1+ 4 \Bar{w}_0^2}},
    \label{mag}
\end{equation}
with $ \Bar{w}_0 = \frac{w(\frac{1}{2},-\frac{1}{2})}{|w(\frac{1}{2},\frac{1}{2})-w(-\frac{1}{2},-\frac{1}{2})|} $, while the spin-spin correlation function between the Ising nodes is
\begin{equation}
    \langle \sigma_i \sigma_{i+r}  \rangle = \langle \sigma^z \rangle + \Big(\frac{\Bar{w}_0}{B}\Big)^2 \Big(\frac{\lambda_-}{\lambda_+}\Big)^r .
    \label{spin-spin}
\end{equation}
Here, $r$ denotes the distance between the sites $i$ and $j=i+r$. $B = \sqrt{\big(w(\frac{1}{2},\frac{1}{2}) - w(\frac{-1}{2},-\frac{1}{2}))^2 + 4 w(\frac{1}{2},-\frac{1}{2}\big)^2 }$, and $\lambda_\pm =\frac{1}{2} \Big[ w(\frac{1}{2},\frac{1}{2}) + w(-\frac{1}{2},-\frac{1}{2}) \pm B \Big]$. At the Heisenberg nodes, the spin correlation function are computed through derivatives of the free energy $\mathcal{F} = -\frac{1}{\beta} \log(\lambda_+)$ as
\begin{align}
\langle S^z \rangle &= \frac{\partial \mathcal{F}}{\partial h},
\end{align}
where $h$ is the external magnetic field acting on the Heisenberg nodes. The spin-spin correlations function between the $(a,b)$ nodes follow as
\begin{equation}
\langle S^{\nu}_{a,i} S^{\nu'}_{b,i} \rangle =
\begin{cases}
\frac{\partial\mathcal{F}}{\partial J_{\nu}}
\quad \textbf{\text{if}} \quad  \nu = \nu', \quad \text{with } \nu = \{x, y, z\},\\
 0  \quad\quad\textbf{\text{if}} \quad \
  \nu \neq \nu', 
\end{cases}
\end{equation}
where $ J_{\nu}, \nu = \{x, y, z\},$ denotes the coupling strengths between the Heisenberg nodes in the $``\nu"$ direction. Finally, the Ising-Heisenberg spin correlation function 
\begin{align}
\langle S^{z}_{a,i}\sigma_{j} \rangle = q_0 \langle \sigma_{j} \rangle + q_{0,1} \langle \sigma_{i} \sigma_{j}  \rangle 
+ \langle \sigma_{i+1} \sigma_{j}  \rangle + q_{1,1} \langle \sigma_{i} \sigma_{i+1} \sigma_{j}  \rangle,
\end{align}
with
\begin{align}
q_0 = -\frac{1}{2\beta} \frac{\partial}{\partial h} \ln f_{eff}, \quad 
q_{1,0} = \frac{1}{2} \frac{\partial h_{eff}}{\partial h}, \quad 
q_{1,1} = \frac{1}{8} \frac{\partial J_{eff}}{\partial h}.
\end{align}
The one-body and two-body correlation functions are given by \Cref{mag,spin-spin}, while the three-body correlation function is written as $\langle \sigma_{i} \sigma_{i+1} \sigma_{j}  \rangle = \langle \sigma_{i} \rangle^3 +  \langle \sigma_{i} \rangle (1-  \langle \sigma_{i} \rangle^2)(e^{(i-j)/\xi}+e^{-1/\xi} (1+ e^{(j-i)/\xi}) )$. $\xi$ is the correlation length of the standard Ising chain given by~\cite{baxter2016exactly}
\begin{equation}
    \xi = \log(\frac{\lambda_+}{\lambda_-}).
\end{equation}
Fig.~\eqref{corr_func_spin_half} and~\eqref{corr_func_spin_one} shows the correlation functions of the ATIH chain, Eq.~\eqref{ham}, for the spin$-(\frac{1}{2}-\frac{1}{2})$ and spin$-(\frac{1}{2}-1)$ case, with respect to the set of parameters $(x,y)$ Eq.~\eqref{phase_diagram_1_parameters} and Eq.~\eqref{phase_diagram_2_parameters}, respectively.
\onecolumngrid
\section{Density matrix\label{dm}}
The full expression of the density matrix is written below
\begin{align}
    \rho_{i,a,b,i+1} &= \frac{1}{d} \Bigg[ \mathbb{1}_{d\times d} + \langle \sigma^z_i\rangle~\sigma^z \otimes\mathbb{1}\otimes\mathbb{1}\otimes\mathbb{1} + \langle \sigma^z_{i+1}\rangle~ \mathbb{1}\otimes\mathbb{1}\otimes\mathbb{1}\otimes \sigma^z
    + \langle S^z_a\rangle~\mathbb{1} \otimes S^z \otimes\mathbb{1}\otimes\mathbb{1} +  \langle S^z_b\rangle~\mathbb{1}\otimes\mathbb{1}\otimes S^z\otimes \mathbb{1} \nonumber \\
    &+ \langle \sigma^z_i\sigma^z_{i+1}\rangle~\sigma^z \otimes \mathbb{1}\otimes\mathbb{1}\otimes\sigma^z
    + \langle  S^z_a  S^z_b\rangle~\mathbb{1}\otimes  S^z \otimes S^z \otimes \mathbb{1}
    + \langle  S^x_a  S^x_b\rangle~\Big(\mathbb{1}\otimes  S^x \otimes S^x \otimes \mathbb{1}~
    +~\mathbb{1}\otimes  S^y \otimes S^y \otimes \mathbb{1}\Big)\nonumber \\
        &+ \langle S^z_a \sigma^z_{i+1}\rangle~\Big(\mathbb{1} \otimes \mathbb{1}\otimes S^z\otimes\sigma^z +\mathbb{1} \otimes S^z\otimes \mathbb{1}\otimes\sigma^z\Big)
       + \langle  \sigma^z_i  S^z_b\rangle~\Big(\sigma^z\otimes S^z \otimes \mathbb{1} \otimes \mathbb{1} + \sigma^z\otimes \mathbb{1} \otimes S^z \otimes \mathbb{1}\Big)\nonumber \\
    &+ \langle S^x_a S^x_b \rangle \langle \sigma^z_{i+1} \rangle~\Big(\mathbb{1}\otimes S^x \otimes S^x \otimes \sigma^z + \mathbb{1}\otimes S^y \otimes S^y \otimes \sigma^z\Big) +\langle S^z_a S^z_b \rangle \langle \sigma^z_{i+1} \rangle~\mathbb{1}\otimes S^z \otimes S^z \otimes \sigma^z   \nonumber \\
    &+\langle S^x_a S^x_b \rangle \langle \sigma^z_i \rangle~\Big( \sigma^z \otimes S^x \otimes S^x \otimes \mathbb{1} + \sigma^z \otimes S^y \otimes S^y \otimes \mathbb{1} \Big) +  \langle S^z_a S^z_b \rangle \langle \sigma^z_i \rangle~\sigma^z \otimes S^z \otimes S^z \otimes \mathbb{1}\\
    &+ \langle \sigma^z_i \sigma^z_{i+1} \rangle \langle S^z_a \rangle~\Big(\sigma^z \otimes S^z \otimes \mathbb{1} \otimes \sigma^z + \sigma^z \otimes \mathbb{1} \otimes S^z \otimes \sigma^z\Big)
    \nonumber \\
    &+ \langle  \sigma^z_i \sigma^z_{i+1}  \rangle \langle  S^x_a S^x_{b}   \rangle  
    ~\Big(\sigma^z\otimes  S^x \otimes S^x \otimes \sigma^z
    + \sigma^z\otimes  S^y \otimes S^y \otimes \sigma^z\Big)
    + \langle  \sigma^z_i \sigma^z_{i+1}  \rangle \langle  S^z_a S^z_{b}   \rangle  
    \sigma^z\otimes  S^z \otimes S^z \otimes \sigma^z
    \nonumber \Bigg].
    \label{full_DM}
\end{align}
\twocolumngrid
\bibliography{bib.bib}
\end{document}